\documentclass[twocolumn]{aastex631}

\usepackage{amsmath}
\usepackage{graphicx}
\def\bea{\begin{eqnarray}}
\def\eea{\end{eqnarray}}
\def\f{\frac}
\def\n{\nonumber}
\usepackage{threeparttable}
\usepackage{appendix}
\usepackage{chngcntr}

\begin{document}

\title{One Fits All: A Unified Synchrotron Model Explains GRBs with FRED-Shape Pulses}

\correspondingauthor{Bin-Bin Zhang, Xiao-Hong Zhao}
\email{bbzhang@nju.edu.cn;zhaoxh@ynao.ac.cn}

\author[0009-0008-2841-3065]{Zhen-yu Yan}
\affiliation{School of Astronomy and Space Science, Nanjing University, Nanjing 210093, China}
\affiliation{Key Laboratory of Modern Astronomy and Astrophysics (Nanjing University), Ministry of Education, China}
\author[0000-0002-5485-5042]{Jun Yang}
\affiliation{School of Astronomy and Space Science, Nanjing University, Nanjing 210093, China}
\affiliation{Key Laboratory of Modern Astronomy and Astrophysics (Nanjing University), Ministry of Education, China}
\author[0000-0003-3659-4800]{Xiao-Hong Zhao}
\affiliation{Yunnan Observatories, Chinese Academy of Sciences, Kunming, China}
\affiliation{Center for Astronomical Mega-Science, Chinese Academy of Sciences, Beijing, China}
\affiliation{Key Laboratory for the Structure and Evolution of Celestial Objects, Chinese Academy of Sciences, Kunming, China}
\author[0000-0002-5485-5042]{Yan-zhi Meng}
\affiliation{School of Astronomy and Space Science, Nanjing University, Nanjing 210093, China}
\affiliation{Key Laboratory of Modern Astronomy and Astrophysics (Nanjing University), Ministry of Education, China}
\author[0000-0003-4111-5958]{Bin-Bin Zhang}
\affiliation{School of Astronomy and Space Science, Nanjing University, Nanjing 210093, China}
\affiliation{Key Laboratory of Modern Astronomy and Astrophysics (Nanjing University), Ministry of Education, China}
\affiliation{Purple Mountain Observatory, Chinese Academy of Sciences, Nanjing 210023, China}

\begin{abstract}
The analysis of gamma-ray burst (GRB) spectra often relies on empirical models lacking a distinct physical explanation. Previous attempts to couple physical models with observed data focus on individual burst studies, fitting models to segmented spectra with independent physical parameters. However, these approaches typically neglect to explain the time evolution of observed spectra. In this study, we propose a novel approach by incorporating the synchrotron radiation model to provide a self-consistent explanation for a selection of single-pulse GRBs. Our study comprehensively tests the synchrotron model under a unified physical condition, such as a single injection event of electrons. By tracing the evolution of cooling electrons in a decaying magnetic field, our model predicts time-dependent observed spectra that align well with the data. Using a single set of physical parameters, our model successfully fits all time-resolved spectra within each burst. Our model suggests that the rising phase of the GRB light curve results from the increasing number of radiating electrons, while the declining phase is attributed to the curvature effect, electron cooling, and the decaying magnetic field. Our model provides a straightforward interpretation of the peak energy's evolution, linked to the decline of the magnetic field and electron cooling due to the expansion of the GRB emission region. Our findings strongly support the notion that spectral and temporal evolution in GRB pulses originates from the expansion of the GRB emission region, with an initial radius of approximately $10^{15}$ cm, and synchrotron radiation as the underlying emission mechanism.
\end{abstract}


\section{Introduction} \label{sec:Introduction}

As the 50-year anniversary of gamma-ray burst (GRB) research approaches, the physical origin of the prompt emission in GRBs is still under debate. The generation of the prompt emission involves two major physical processes, namely the energy dissipation of the relativistic ejecta and the radiation mechanism to produce the $\gamma$-ray photons. The most straightforward energy dissipation process invokes the process that gamma-ray photons escape from the ejecta when its opacity is below unity, giving rise to what is known as photospheric emission \citep{1986ApJ...308L..47G,1986ApJ...308L..43P,1991ApJ...369..175A,2000ApJ...530..292M,2017IJMPD..2630018P}. However, the spectra of the photospheric emission are generally thermal or quasi-thermal, which contradicts the non-thermal spectrum observed in most GRB instances. This implies that the photospheric emission is not the dominant dissipation mechanism, unless additional factors or mechanisms are considered \citep[e.g.][]{2000ApJ...530..292M,2005ApJ...628..847R,2010MNRAS.407.1033B,2011ApJ...732...49P,2013MNRAS.428.2430L,2018ApJ...860...72M,2022ApJS..263...39M}.

Among the most widely recognized dissipation models are the internal shock model \citep[e.g.,][]{1994ApJ...427..708P,1994ApJ...430L..93R,1998MNRAS.296..275D}, based on the matter-dominated fireball, and the magnetic dissipation model, based on the Poynting flux-dominated outflow \citep{2003astro.ph.12347L,2011ApJ...726...90Z}. In the internal shock model, a fraction of the kinetic energy carried by the GRB ejecta is converted into internal energy, which is then radiated through non-thermal mechanisms, such as synchrotron emission or IC scattering. This model can also account for the observed light curves \citep[e.g.,][]{1997ApJ...490...92K}. However, the internal shock model faces several significant challenges, such as the so-called fast cooling problem \citep{2000MNRAS.313L...1G} and the low radiative efficiency problem \citep{1999ApJ...523L.113K,1999ApJ...522L.105P}. To overcome the efficiency problem, extreme conditions are required \citep{2000ApJ...539L..25B,2001ApJ...551..934K}. Alternatively, magnetic dissipation models, such as the Internal-Collision-Induced MAgnetic Reconnection and Turbulence (ICMART) model \citep{2011ApJ...726...90Z}, have been proposed and extensively investigated, providing a solution to the issues encountered by the internal shock model. In the ICMART model, the magnetic energy is converted into the energy of electrons by the magnetic reconnection mechanism. GRBs are further produced by the synchrotron radiation of the accelerated electrons. This model requires larger radii and a weaker magnetic field compared to the internal shock model, with the magnetic field strength decreasing as the radius increases. 

In both dissipation models, synchrotron radiation is the most natural and efficient mechanism to produce GRBs. Electrons are accelerated either within internal shocks or in the magnetic reconnection region, ultimately emitting gamma rays through the synchrotron process. Extensive research has specifically focused on the synchrotron model, exploring its ability to account for both the spectral characteristics and temporal profiles of GRBs \citep{1998MNRAS.296..275D,2011A&A...526A.110D}. To address the challenge of the fast cooling problem, additional effects like inverse Compton (IC) cooling, the Klein--Nishina effect, magnetic field decay, and adiabatic cooling have been explored to comprehensively explain the observed GRB spectrum \citep{2006ApJ...653..454P,2009ApJ...705.1714A,2009ApJ...698L..98W,2011A&A...526A.110D,2014NatPh..10..351U,2014ApJ...780...12Z,2018ApJS..234....3G}. Another problem of the synchrotron model in interpreting the GRB spectra is that the low energy index in a part of bursts exceeds 1/3 ($F_\nu$ spectrum), resulting in the so-called ``synchrotron line of death” problem \citep{1998ApJ...506L..23P}. 
However, recent studies have revealed that the synchrotron model aptly conforms to the time-resolved gamma-ray spectra 
\citep[e.g.,][]{2011ApJ...741...24B,2014ApJ...784...17B,2014ApJ...784L..43B,Zhang_2016,2018NatAs...2...69Z,2018ApJ...867...52X,2020NatAs...4..174B,2021ApJ...920...53C,2022ApJ...926..178W}. The broadband spectra, encompassing both X-ray and optical bands, also strongly align with the synchrotron model predictions \citep{2017ApJ...846..137O,2018A&A...616A.138O,2018A&A...613A..16R}.

A limitation of previous time-resolved gamma-ray spectral fitting approaches is that they are often applied independently to each time segment, without considering the evolution of emissions in adjacent intervals. Consequently, the obtained best-fit parameters for individual time intervals lack a coherent initial physical condition and may not align with the overall evolution of physical parameters predicted by the magnetic dissipation model. This challenge becomes particularly pronounced when dealing with GRBs that exhibit complex light-curve profiles with overlapping pulses, as these characteristics can be attributed to the intrinsic behaviors of the central engines. Nevertheless, a prior investigation conducted by \cite{2023ApJ...947L..11Y} presented a compelling case study wherein the synchrotron radiation model, incorporating an expanding emission region with relativistic speed and a global magnetic field that decays with radius, effectively reproduced the complete time series of precursor spectra for the Brightest of All-Time (``BOAT") GRB 221009A. The precursor of GRB 221009A exhibits a simple Fast-Rise and Exponential-Decay \citep[i.e., FRED;][]{1996ApJ...473..998F} shape, suggesting a straightforward evolving emission unit. It is intriguing to explore whether the novel approach proposed by \cite{2023ApJ...947L..11Y} implies that this synchrotron model can be applied to other typical GRBs characterized by simple shapes, such as FRED profiles. This objective serves as the focal point of this study. 

In this study, we first give a detailed description of our synchrotron model in Section \ref{sec:Model}. Then we collect a sample of GRBs comprising FRED pulses with known redshift, utilizing the data obtained from the {\it Fermi} Gamma-ray Burst Monitor (GBM) instrument mission spanning the period from 2008 to 2022, predating GRB 221009A. The processes of data reduction and sample selection are described in Section \ref{sec:DATA}. After conducting comprehensive data analyses, we apply the synchrotron model to perform spectral fits for these GRBs. The results of our analyses are presented in Section \ref{sec:TheFit}, followed by discussions in Section \ref{sec:Discussion} and a brief summary in Section \ref{sec:Conclusions}.

\section{The Synchrotron Model} \label{sec:Model}
The physical model applied in this study to fit the spectral data of GRBs is a synchrotron model that involves a spherical thin shell expanding relativistically with a constant Lorentz factor of $\Gamma=1/\sqrt{1-{\beta}^2}$, where $\beta$ is the dimensionless speed of the shell. Electrons in the shell are accelerated to high energies by various mechanisms, such as magnetic reconnection (e.g., \citealt{2011ApJ...726...90Z}). The accelerated electrons in the magnetic field of the shell produce a GRB by synchrotron radiation mechanism and subsequently cool down to low energies due to radiation loss. Additionally, the cooling process is further influenced by the adiabatic expansion of the shell and the IC cooling. 

To track the evolution of electron distribution, we only follow the cooling process while assuming that the acceleration of electrons takes place on a much shorter time scale. Using the methodology outlined in Appendices \ref{sec:CIP} and \ref{sec:B}, we solve the continuity equation of electrons in the co-moving frame of the shell, which can be expressed as
\begin{equation} \label{eq:electron}
\frac{\partial}{\partial R} \left(\frac{{\rm d}N_{\rm e}^{\prime}}{{\rm d}\gamma^{\prime}_{\rm e}} \right) +
\frac{\partial}{\partial \gamma^{\prime}_{\rm e}} \left[ \frac{{\rm d} \gamma^{\prime}_{\rm e}}{{\rm d} R} \left(\frac{{\rm d}N_{\rm e}^{\prime}}{{\rm d}\gamma^{\prime}_{\rm e}} \right) \right]=Q'\left( \gamma^{\prime}_{\rm e} \right),
\end{equation}
where ${\rm d}N_{\rm e}^{\prime}/{\rm d}\gamma^{\prime}_{\rm e}$ is the number of electrons per electron energy interval, $Q'\left( \gamma^{\prime}_{\rm e}\right)$ is the electron injection rate with units of ${\rm cm}^{-1}$, $R=c\beta\Gamma (t'_0+t')$ is the source-frame radius, and $t^\prime$ is the elapsed time since an initial time $t^\prime_0$. In this context, all quantities with a prime are measured in the co-moving frame. ${\rm d} \gamma^{\prime}_{\rm e}/{\rm d} R$ is the cooling rate of the electrons and is given by
\begin{equation} \label{eq:cooling}
\frac{{\rm d} \gamma^{\prime}_{\rm e}}{{\rm d} R}= \left(\frac{{\rm d} \gamma^{\prime}_{\rm e}}{{\rm d} R}\right)_{\rm syn} + \left(\frac{{\rm d} \gamma^{\prime}_{\rm e}}{{\rm d} R}\right)_{\rm adi}=
-\frac{\sigma_{\rm T} {B^{\prime}}^2 {\gamma^{\prime}_{\rm e}}^2}{6 {\rm \pi} m_{\rm e} c^2 \beta \Gamma}
-\frac{2\gamma^{\prime}_{\rm e}}{3R},
\end{equation}
where $\left({\rm d} \gamma^{\prime}_{\rm e}/{\rm d} R\right)_{\rm syn}$ is the synchrotron radiation cooling rate, $\left({\rm d} \gamma^{\prime}_{\rm e}/{\rm d} R\right)_{\rm adi}$ is the adiabatic cooling rate, $\sigma_{\rm T}$ is the Thomson cross section, $B^{\prime}$ is the magnetic field strength in the shell, $m_{\rm e}$ is the electron mass, and $c$ is the speed of light. The synchrotron radiation cooling rate can dominate early if the initial magnetic field strength is high enough, while at late times, the adiabatic cooling rate can dominate over the synchrotron cooling rate because of the decaying of magnetic field strength. In our study, we neglect the IC cooling due to the Klein--Nishina effect, as the condition $\gamma_{\rm m}^{\prime} E_{\rm p}/\Gamma>m_{\rm e} c^2$ is generally satisfied for the global magnetic field configuration. Here, $E_{\rm p}$ denotes the spectral peak in the energy spectrum, and $\gamma_{\rm m}^{\prime}$ is the minimum electron injection energy. 

We assume that electrons with a power-law distribution between $\gamma ^{\prime} _{\rm m}$ and $\gamma ^{\prime} _{\rm e,max}$ are uniformly and steadily injected into the shell before an injection stop time $t_{\rm inj}$ measured in the observer frame. When the condition $t^{\prime}<2\Gamma t_{\rm inj}/(1+z)$ is satisfied, where $z$ is the redshift, the injection rate is given by
\begin{equation} \label{eq:injection}
Q'\left( \gamma^{\prime}_{\rm e}\right)=
\begin{cases}
Q_{\rm 0} {\gamma^{\prime}_{\rm e}}^{-p}, & \gamma ^{\prime} _{\rm m} < \gamma^{\prime}_{\rm e} < \gamma^{\prime}_{\rm e,max}\\
0, & \text{otherwise}
\end{cases},
\end{equation}
where $Q_{\rm 0}$ is the injection coefficient in ${\rm cm^{-1}}$, and $p$ is the power-law index of the injected electron distribution. To ensure simplicity and avoid possible mathematical, yet nonphysical, degeneracy, we fix the value of $\gamma^{\prime}_{\rm e,max}$ at $10^8$.

The magnetic field in Equation \eqref{eq:cooling} decays with time due to the expansion of the emission shell. Nevertheless, the specific functional form describing the magnetic field decay as the shell radius increases remains not well studied. In accordance with \cite{2014NatPh..10..351U}, we adopt a power-law decay model for the magnetic field based on the principle of magnetic flux conservation \citep{2001A&A...369..694S}. Such a power-law decay is expressed as follows:
\begin{equation} \label{eq:magnetic}
B^{\prime}=B^{\prime}_{\rm 0} {\left( \frac{R}{R_{\rm 0}} \right)}^{-\alpha_B},
\end{equation}
where $R_{\rm 0}= c \beta \Gamma t'_0$ represents the initial radius at which electrons are injected, $B'_0$ denotes the magnetic field strength at the initial radius, and $B'$ corresponds to the magnetic field strength at $R$.

With the specified assumptions detailed in Equations \eqref{eq:cooling}, \eqref{eq:injection} and \eqref{eq:magnetic}, we adopt two numerical schemes described in Appendices \ref{sec:CIP} and \ref{sec:B} to numerically solve Equation \eqref{eq:electron} and derive the electron distribution at any $R$ (or any source-frame time $t_{\rm e}=R/\beta c$). The synchrotron spectral power at a given frequency $\nu^{\prime}$ for a given electron distribution can be expressed as \citep{1979rpa..book.....R}:
\begin{equation} \label{eq:power}
P^{\prime}\left(\nu^{\prime}\right)=\frac{\sqrt{3} {q_{\rm e}}^{3} B^{\prime}}{m_{\rm e} c^{2}} \int_{\gamma^{\prime}_{\rm e,min}}^{\gamma^{\prime}_{\rm e,max}} \left(\frac{{\rm d} N_{\rm e}^{\prime}}{{\rm d} \gamma^{\prime}_{\rm e}}\right) F\left(\frac{\nu^{\prime}}{\nu_{\rm c}^{\prime}}\right) {\rm d} \gamma^{\prime}_{\rm e},
\end{equation}
where $\nu_{\rm c}^{\prime}=3 q_{\rm e} B^{\prime} {\gamma^{\prime}_{\rm e}}^2 /\left(4 \pi m_{\rm e} c\right)$, $q_{\rm e}$ is the electron charge, $F(x)=x \int_{x}^{+\infty} K_{\rm 5 / 3}(k) d k$, $ K_{\rm 5 / 3}(k)$ is the Bessel function, and $\gamma^{\prime}_{\rm e,min}$ is the minimum electron energy used in the calculation, which is distinct from the minimum injection energy $\gamma ^{\prime}_{\rm m}$. 

Next, considering the spherical geometry of the shell and the equal-arrival-time surface (EATS) effect, we can obtain the observed flux \citep{ 2018ApJS..234....3G} as follows:
\begin{equation} \label{eq:Fv_theta}
F_{\nu_{\rm obs}}=\frac{1+z}{4 \pi {D_{\rm L}}^{2}} \int_{\rm 0}^{\theta_{\rm max}} 
P^{\prime}\left(\nu^{\prime}\left(\nu_{\rm obs}\right)\right)
\mathcal{D}^3 \frac{{\rm sin} \theta}{2} {\rm d} \theta,
\end{equation}
where $\theta$ is the angle formed between the line of sight (LOS) and the velocity of a point on the shell, $\mathcal{D}=1/[\Gamma (1-\beta \cos\theta)]$ is the Doppler factor, $\nu _{\rm obs} = \nu^{\prime} \mathcal{D}/(1+z)$ is the observed photon frequency, $\theta_{\rm max}$ is the largest angle on the EATS, and $D_{\rm L}$ is the luminosity distance calculated by adopting a flat $\Lambda$CDM universe using the cosmological parameters $H_{\rm 0} =67.4 \ \rm{km }\, \rm{s} ^{-1} \, \rm{Mpc}^{-1}$, $\Omega _{\rm m} =0.315$ and $\Omega _{\rm \Lambda} =0.685$ \citep{2020A&A...641A...6P}. The $\frac{1}{2}$ factor in Equation \eqref{eq:Fv_theta} arises from the ratio of the area of the ring ($2\pi R^2\sin \theta \mathrm{d} \theta$) to the area of the entire sphere ($4\pi R^2$) in the EATS integral.

Assuming that the electron injection starts at $t_{\rm s}=0$ s in the observer frame, the received flux by the observer at a given observer-frame time $t_{\rm{obs}}$ can be calculated as the integral of emission over the EATS using Equation \eqref{eq:EATS},
\begin{equation} \label{eq:EATS}
R=\frac{\beta c (t_{\rm obs,0}+t_{\rm obs})}{(1+z)(1-\beta \cos\theta)},
\end{equation}
where $t_{\rm obs,0}=R_0(1-\beta)(1+z)/(\beta c )$.
Rewriting this equation as 
\begin{equation} \label{eq:theta}
\theta=\arccos{\left[ \frac{1}{\beta} - \frac{c (t_{\rm obs,0}+t_{\rm obs})}{(1+z)R} \right] },
\end{equation}
we can find that for a given $t_{\rm{obs}}$, $\theta$ reaches its maximum value, $\theta_{\rm max}$, when $R =R_0$, namely,
\begin{equation} \label{eq:theta_max}
\theta_{\rm max}=\theta_{R_0}=\arccos{\left[ 1-\frac{ct_{\rm obs}}{(1+z)R_0}\right]}. 
\end{equation}
Noticeably, if $\theta_{R_0}$ exceeds the half-opening angle $\theta_{\rm j}$ of a jet, a portion of the EATS will lie beyond the jet edge. Therefore, the upper limit of the integral in Equation \eqref{eq:Fv_theta} should be the minimum angle between $\theta_{R_0}$ and $\theta_{\rm j}$, i.e., $\theta_{\rm max}= \min (\theta_{R_0}, \theta_{\rm j})$. 
Taking a typical parameter set as $\Gamma=300$, $z=1$, $R_{\rm 0}=10^{15}$ cm, and $t_{\rm obs}=30$ s, we obtained $\theta_{R_0} \approx 0.03 ~ {\rm rad}$ from Equation \eqref{eq:theta_max}. This value is much lower than the half-opening angles of GRB jets with a typical value of $\theta_{\rm j} \sim 0.1$ rad. Therefore, we take the value of $\theta_{R_0}$ for $\theta_{\rm max}$ while assuming that the effect of the jet edge is negligible in this study.

For a given $t_{\rm obs}$, $\Gamma$ and $z$, Equation \eqref{eq:EATS} can be differentiated in the following format:
\begin{equation} \label{eq:EATS_differentiated}
\frac{c}{R}(1-\beta \cos \theta){\rm d}t_{\rm e}=-{\rm sin} \theta {\rm d} \theta.
\end{equation}
Using ${\rm d}t_{\rm e}$ in Equation \eqref{eq:EATS_differentiated} to replace ${\rm d}\theta$ in Equation \eqref{eq:Fv_theta}, one can obtain
\begin{equation} \label{eq:Fv_t}
F_{\nu_{\rm obs}}=\frac{c(1+z)}{8 \pi {D_{\rm L}}^{2}} \int_{t_{\rm e,0}}^{t_{\rm e,max}} 
\frac{P^{\prime}\left(\nu^{\prime}\left(\nu_{\mathrm{obs}}\right)\right)}{R{\Gamma}^3 {\left( 1-\beta \cos\theta\right)}^2} {\rm d} t_{\rm e},
\end{equation}
where $t_{\rm e,0}=R_{\rm 0}/\beta c$ is the source-frame initial time, and $t_{\rm e,max}=t_{\rm e,0} +t_{\rm obs}/(1+z)(1-\beta)$ is the maximum source-frame time for the integral of the EATS at $t_{\rm obs}$. We note that an independent derivation of Equation \eqref{eq:Fv_t} has been conducted by \cite{2015ApJ...808...33U}. 

Combining Equations \eqref{eq:power} and \eqref{eq:Fv_t}, the observed flux can be expressed as
\begin{eqnarray} \label{eq:Fv_total}
&&F_{\nu_{\rm obs}} = F_{\nu_{\rm obs}}(t_{\rm obs},\nu_{\rm obs}, B^{\prime}_{\rm 0}, \alpha_B, \gamma^{\prime}_{\rm m}, \Gamma, p, t_{\rm inj}, R_{\rm 0}, Q_{\rm 0}) \nonumber \\
&& =C_0 \frac{ B^{\prime}_{\rm 0} {R_{\rm 0}}^{\alpha_B}} {c^{\alpha_B }{\Gamma}^3 \beta^{\alpha_B +1}} \times \\
&& \int_{t_{\rm e,0}}^{t_{\rm e,max}} 
 \frac{{t_{\rm e}}^{-\alpha_B-1}} {{\left( 1-\beta \cos\theta\right)}^2} 
 \int_{\gamma^{\prime}_{\rm e,min}}^{\gamma^{\prime}_{\rm e,max}} 
 \left(\frac{{\rm d} N_{\rm e}^{\prime}}{{\rm d} \gamma^{\prime}_{\rm e}}\right) 
 F\left(\frac{\nu^{\prime} }{\nu_{\rm c}^{\prime}}\right) {\rm d} \gamma^{\prime}_{\rm e} \;\; {\rm d} t_{\rm e}, \nonumber 
\end{eqnarray}
where $C_0=\sqrt{3} {q_{\rm e}}^3 (1+z)/ ({8\pi m_{\rm e} c^{2} {D_{\rm L}}^2})$. 

In summary, for a known redshift $z$, the electron energy distribution ${\rm d}N_{\rm e}^{\prime}/{\rm d}\gamma^{\prime}_{\rm e}$ can be derived by solving Equation \eqref{eq:electron} using a given set of parameters, $\mathcal{P}=(B^{\prime}_{\rm 0}, \alpha_B, \gamma^{\prime}_{\rm m}, \Gamma, p, t_{\rm inj}, R_{\rm 0}, Q_{\rm 0})$. Subsequently, by evaluating the double integral in Equation \eqref{eq:Fv_total}, the flux at any observer-frame time $t_{\rm obs}$ and any observed photon frequency $\nu_{\rm obs}$ is expressed as follows:
\begin{equation} \label{eq:Fv_short}
F_{\nu_{\rm obs}}=F_{\nu_{\rm obs}}(t_{\rm obs},\nu_{\rm obs}, \mathcal{P}).
\end{equation}

We note, however, that obtaining $F_{\nu_{\rm obs}}$ in Equation \eqref{eq:Fv_short} and comparing it to the observed data (Section \ref{sec:TheFit}) can be computationally costly, as it involves solving Equation \eqref{eq:electron} numerically, as well as a double integral in Equation \eqref{eq:Fv_total}. Therefore, this study requires a high-performance computational resources.

\section{Data Reduction and Sample Selection} \label{sec:DATA}

To undertake a comparative analysis between our model's predictions, as given by Equation \eqref{eq:Fv_short}, and the observed data, it is imperative to carefully select a well-established sample comprising bursts whose light curves exhibit the FRED shape. This distinctive shape, as extensively elucidated in the comprehensive analysis conducted by \cite{2023ApJ...947L..11Y}, facilitates the investigation of the burst's time-resolved spectra within the context of the synchrotron emission model as elaborated in Section \ref{sec:Model}. The {\it Fermi}/GBM instrument is particularly well suited for our time-resolved spectral analysis due to its wide spectral coverage spanning from 8 keV to 40 MeV and high temporal resolution in microseconds. The data reduction and analysis procedures in this study closely align with the methodology outlined by \cite{2011ApJ...730..141Z,Zhang_2016,2020ApJ...899..106Y,Yang2022Natur,2023ApJ...947L..11Y}. Specifically, our sample selection, data reduction, and data analysis procedures adhere to the following sequential steps.

\begin{enumerate}
 \item Redshift screening. The redshift of a GRB is crucial to determine its luminosity and intrinsic time scale of the light curve. Furthermore, the acquisition of the redshift information is imperative for constraining essential parameters of the GRB ejecta, such as its radius, bulk Lorentz factor, and injection rate, through the utilization of the synchrotron model fitting. Therefore, we have collected GRB data observed by {\it Fermi}/GBM with known redshift. We refer to the {\it Fermi} GBM Burst Catalog\footnote{\url{https://heasarc.gsfc.nasa.gov/W3Browse/fermi/fermigbrst.html}} \citep{,2014ApJS..211...12G,2014ApJS..211...13V,2020ApJ...893...46V,2016ApJS..223...28N}, which contains a total of 3377 GRBs detected by {\it Fermi}/GBM detectors predating GRB 221009A. From this dataset, we identify 209 GRBs with measured redshift values by cross-referencing the ``Big Table"\footnote{\url{https://www.mpe.mpg.de/~jcg/grbgen.html}} and the ``GRBweb"\footnote{\url{https://user-web.icecube.wisc.edu/~grbweb_public/Summary_table.html}} of well-localized GRBs.
 
 \item Light-curve screening. To achieve a well-fitted set of time-resolved spectra, we carefully curate the sample by exclusively selecting the brightest bursts exhibiting a FRED shape while ensuring a substantial signal-to-noise ratio (S/N) among the 209 bursts. To do so, we firstly retrieve the time-tagged event data set from the {\it Fermi}/GBM public data archive\footnote{\url{https://heasarc.gsfc.nasa.gov/FTP/fermi/data/gbm/daily/}} for each of the 209 GRBs, bin the photon count events for each {\it Fermi}/GBM detector, and derive the 10--1000 keV light curves with a bin size of 0.5 s. Secondly, for the sake of better subtracting the background, we employ {\it numpy.polyfit}\footnote{\url{https://numpy.org/doc/stable/reference/generated/numpy.polyfit.html}} to obtain the background curve of the GRB using the interpolation of the second-order polynomial fitting result. The uncertainties of the background and net photon counts are estimated using {\it pwkit.kbn\_conf}\footnote{\url{https://pwkit.readthedocs.io/en/latest/_modules/pwkit/kbn_conf/}}. Then, we use {\it gv\_significance}\footnote{\url{https://github.com/giacomov/gv_significance}} \citep{2018ApJS..236...17V} to calculate the S/N. At last, we limit the S/N to S/N $>$ 8 and select 129 bright bursts.

 Furthermore, in order to select the GRBs with light curves exhibiting FRED shapes or containing FRED pulses, we thoroughly examine the overall shape of each pulse among the selected 129 GRBs. We first use {\it numpy.histogram\_bin\_edges}\footnote{\url{https://numpy.org/doc/stable/reference/generated/numpy.histogram_bin_edges.html}} to automatically bin the photon count events from the detector with the highest S/N. Then we fit the light curve with a functional pulse model introduced by \cite{2005ApJ...627..324N} using {\it emcee} \citep{2013PASP..125..306F},
\begin{equation} \label{eq:Norris}
I(t)=A \exp{[2\mu -\tau_1 /(t-t_{\rm 0}) - (t-t_{\rm 0})/\tau_2]} ,
\end{equation}
 where $A$ is the pulse amplitude, $t_{\rm 0}$ is the pulse start time, $t$ is the time since $t_{\rm 0}$, $\tau_1$ and $\tau_2$ are two free parameters altering the pulse profile, and $\mu=(\tau_1/\tau_2)^{1/2}$. According to the fitting goodness, we further screen the sample by only selecting 16 well-fitted isolated pulses with $\chi^2/{\rm dof}<1.5$ from the 129 GRBs. It has been verified that the 16 pulses are well separated from other emission (if it exists). The separations, $t_{\rm int}$, defined as the time intervals between the moments when pulses emerge from the background and other emissions start to rise from the background, are all greater than one and a half times the pulse widths $w$, i.e., $t_{\rm int}>1.5w$, where the pulse widths are defined as the time intervals between the two $1/{\rm e}$ intensity points of the pulses, $w=\Delta \tau_{1/{\rm e}}=\tau_2(1+4\mu)^{1/2}$. 
 Next, in order to select the pulses with FRED profiles from the 16 single pulses, we use the best-fit parameters to calculate and compare the rise time $\tau_{\rm rise}=0.5\tau_2\left[ (1+4\mu)^{1/2}- 1\right]$ and the decay time $\tau_{\rm dec}=0.5\tau_2\left[ (1+4\mu)^{1/2}+ 1\right]$. We only select 13 pulses that satisfy $\tau_{\rm dec} > 1.5 ~\tau_{\rm rise}$, since the FRED-shaped light curve necessitates that the duration of the decay phase surpasses that of the rise phase. The final sample number is 13 after the light-curve screening.
 \item Spectral screening. The synchrotron model under consideration in this study is only capable of producing a hard-to-soft evolution pattern for $E_{\rm p}$, the details of which will be discussed in Section \ref{sec:DiscussionEp}. Therefore, we further screen the sample by selecting the pulses that exhibit this $E_{\rm p}$ evolution pattern among the 13 FRED-shaped pulses. First of all, we need to obtain the spectral properties of the 13 pulses observed by {\it Fermi}/GBM.
 {\it Fermi}/GBM comprises a total of 14 detectors, each with a unique orientation, allowing multiple detectors to detect a single GRB simultaneously. Therefore, we select 2--4 detectors with the optimal viewing angles for each GRB and segment the time interval into 5--10 time slices based on the burst brightness as discussed in \cite{2023ApJ...947L..11Y}. For each time slice, we obtain the observed source spectrum of each selected detector by summing up the number of total photons for each energy channel. The corresponding background spectrum is determined by simulating the background level using the baseline algorithm\footnote{\url{https://github.com/derb12/pybaselines}} for each energy channel. Moreover, the detector response matrix is generated using the {\it gbm\_drm\_gen}\footnote{\url{https://github.com/grburgess/gbm_drm_gen}} package \citep{Burgess2018MNRAS, Berlato2019ApJ}. The spectral data will also be utilized for fitting physical models in Section \ref{sec:TheFit}. To determine the spectral evolution pattern, we perform cutoff power-law (CPL) fitting to the spectrum of each time slice. According to the CPL fitting results, we finally select eight pulses with the hard-to-soft $E_{\rm p}$ evolution pattern. 
\end{enumerate}
\begin{table*} 
\centering
\small
\addtolength{\leftskip} {-1.95cm}
\caption{Observed Characteristics of Selected GRB Pulses}
\label{table:GRBs}
\begin{threeparttable}
\begin{tabular}{lccccc}
\toprule
GRB & $z$ & $T_{90}/(1+z)$ (s) & $E_{\rm p}(1+z)$ (keV) & $E_{\rm \gamma,iso}$ ($10^{51}$erg) & Ref. \\ \hline
120326A & 1.798 & 4.2 & $112.8\pm 10.9$ & $31.5\pm 1.2$ & \cite{2015ApJ...814....1L}\\
131011A & 1.874 & 26.8 & $788\pm 71$ & $83.4\pm 0.6$ & \cite{2020MNRAS.492.1919M}\\
140606B & 0.384 & 4.9 & $352_{-37}^{+46}$ & $2.5\pm 0.2$ & \cite{2020MNRAS.492.1919M}\\
150514A & 0.807 & 1.09 & $108_{-13}^{+11}$ & $8.78_{-0.78}^{+0.78} $ & \cite{2020MNRAS.492.1919M}\\
151027A pulse 1 & 0.81 & 2.83 & $302.7_{-32.4}^{+37.1}$ & $7.52_{-0.43}^{+0.50}$ & \cite{2020MNRAS.492.1919M}\\
170607A & 0.557 & 14.5 & $226\pm 18$ & $9.15\pm 0.05$ & \cite{2020MNRAS.492.1919M}\\
190829A pulse 1 & 0.0785 & 5.84 &$129.4 _{-41.1}^{+120.5}$ & 0.032 &\cite{2020ApJ...898...42C}\\
210204A pulse 2 & 0.876 & 11.88 & $370 \pm 56$ & 12.2 &\cite{2022MNRAS.513.2777K}\\
\hline
\end{tabular}
\begin{tablenotes}
\addtolength{\leftskip} {1.4cm}
\footnotesize
\item {\it Note}. 
Multiple episodes are present in the three GRBs with pulse numbers in the first column. For the three cases, our selection solely pertains to the first peak in the initial episode of GRB 151027A, the first episode of GRB 190829A, and the second episode of GRB 210204A. The pulse numbers are omitted from subsequent tables and figures for the sake of simplicity. $z$: redshift; $T_{90}$: time
intervals from 5\% to 95\% of the total fluence;
$E_{\rm p}$: peak energy in the energy spectrum; $E_{\rm \gamma,iso}$: isotropic bolometric emission energy. 
In particular, for the first peak of GRB 151027A, $T_{90}$ in 10--1000 keV, $E_{\rm \gamma,iso}$ in 1--10000 keV, and $E_{\rm p}$ are calculated by this work. The spectral properties are obtained using the results of CPL fits to the time-integrated spectra of b1, n0, n1, and n3 detectors of {\it Fermi}/GBM from $-0.5$ s to $6.2$ s. The error bars represent the $1\sigma$ uncertainties.
\end{tablenotes}
\end{threeparttable}
\end{table*}

\begin{table*}
\addtolength{\leftskip} {-3.2cm}
\caption{Fitting Results}
\label{table:fit}
\begin{threeparttable}
\begin{tabular}{cccccccccccc}
\toprule
GRB & $\mathrm{log} [B^{\prime}_{\rm 0}(\mathrm{G})]$ & $\alpha_B$ & $\mathrm{log} \gamma^{\prime}_{\rm m}$ & $\mathrm{log} \Gamma$ & $p$ & $t_\mathrm{inj}(\mathrm{s})$ & $\mathrm{log} [R_\mathrm{0}(\mathrm{cm})]$ & $\mathrm{log} [Q_0({\rm cm}^{-1})] $ & PGSTAT/DOF \\ \hline

120326A & $1.24_{-0.19} ^{+0.14}$ & $1.67_{-1.10}^{+0.01}$ & $4.86_{-0.12} ^{+0.13}$ & $2.14_{-0.49} ^{+0.29}$ & $4.53_{-0.85} ^{+0.15}$ & $3.30_{-0.22} ^{+0.47}$ & $15.55_{-1.03} ^{+0.54}$ & $54.12_{-3.95} ^{+0.65}$ &774.03/1217 \\
131011A & $-0.05_{-0.01} ^{+0.42}$ & $1.61_{-0.52}^{+0.19}$ & $5.79_{-0.24} ^{+0.02}$ & $2.03_{-0.24} ^{+0.53}$ & $3.77_{-0.24} ^{+0.44}$ & $3.80_{-0.02} ^{+2.43}$ & $15.21_{-0.58} ^{+0.96}$ & $53.62_{-2.25} ^{+1.30}$ &1738.90/2890 \\
140606B & $1.36_{-0.26} ^{+0.11}$ & $1.93_{-0.29} ^{+0.003}$ & $5.37_{-0.22} ^{+0.08}$ & $1.90_{-0.25} ^{+0.76}$ & $4.36_{-0.65} ^{+0.40}$ & $2.05_{-0.21} ^{+0.13}$ & $14.62_{-0.51} ^{+1.51}$ & $54.19_{-4.40} ^{+0.73}$ & 1835.72/2908 \\
150514A & $0.99_{-0.63} ^{+0.05}$ & $1.31_{-0.08} ^{+0.26}$ & $4.87_{-0.03} ^{+0.22}$ & $2.06_{-0.34} ^{+0.27}$ & $5.37_{-0.16} ^{+0.16}$ & $0.81_{-0.05} ^{+0.04}$ & $14.87_{-0.67} ^{+0.55}$ & $58.61_{-0.45} ^{+1.67}$ &3087.09/4872 \\
151027A & $0.14_{-0.29} ^{+0.18}$ & $1.34_{-0.13} ^{+0.46}$ & $5.22_{-0.05} ^{+0.001}$ & $2.52_{-0.12} ^{+0.29}$ & $4.51_{-0.25} ^{+0.07}$ & $0.46_{-0.01} ^{+0.28}$ & $15.62_{-0.23} ^{+0.56}$ & $54.62_{-1.23} ^{+0.05}$ & 1589.32/2427 \\
170607A & $3.41_{-0.96} ^{+0.10}$ & $1.37_{-0.27} ^{+0.23}$ & $4.17_{-0.08} ^{+0.44}$ & $1.46_{-0.15} ^{+0.42}$ & $5.25_{-0.32} ^{+0.30}$ & $2.65_{-0.10} ^{+0.11}$ & $14.32_{-0.31} ^{+0.86}$ & $56.67_{-1.81} ^{+1.98}$ &1134.98/1817 \\
190829A & $1.05_{-0.17} ^{+0.24}$ & $1.74_{-0.36} ^{+0.15}$ & $4.93_{-0.14} ^{+0.11}$ & $1.85_{-0.42} ^{+0.24}$ & $4.08_{-0.38} ^{+0.55}$ & $2.71_{-0.12} ^{+3.93}$ & $14.66_{-0.88} ^{+0.49}$ & $49.92_{-2.06} ^{+3.48}$ &1315.04/2200 \\
210204A & $1.87_{-0.13} ^{+0.82}$ & $1.89_{-0.36} ^{+0.01}$ & $5.11_{-0.25} ^{+0.21}$ & $2.39_{-0.57} ^{+0.75}$ & $3.02_{-0.41} ^{+0.30}$ & $5.89_{-1.16} ^{+1.26}$ & $14.97_{-1.35} ^{+0.95}$ & $45.25_{-3.27} ^{+2.74}$ & 1167.92/1802 \\

\hline
\end{tabular}
\end{threeparttable}
\end{table*}

The final selection consists of eight distinct individual pulses, each chosen from separate GRBs as detailed in Table \ref{table:GRBs}, namely, GRB 120326A, GRB 131011A, GRB 140606B, GRB 150514A, GRB 151027A, GRB 170607A, GRB 190829A, and GRB 210204A. For each pulse, the selected detectors and segmented time slices of the final sample are in Appendix \ref{sec:C}. In particular, for the instances where multiple episodes of GRBs are present in the sample, it is imperative to emphasize that our selection solely pertains to the first peak in the initial episode of GRB 151027A, the first episode of GRB 190829A, and the second episode of GRB 210204A. Table \ref{table:GRBs} provides some key properties for these GRBs, including redshift $z$, rest-frame duration $T_{90}/(1+z)$, rest-frame peak energy $E_{\rm p}(1+z)$, and isotropic energy $E_{\gamma,\rm iso}$.

\section{The Fit} \label{sec:TheFit}
In order to fit the time-resolved spectra in our sample with the time-dependent synchrotron model described in Section \ref{sec:Model}, we employ the Python package {\it MySpecFit} \citep{Yang2022Natur, 2023ApJ...947L..11Y}, which wraps nested sampling implementation {\it Multinest} \citep{Feroz2008MNRAS, Feroz2009MNRAS, Buchner2014A&A, Feroz2019OJAp} to perform the Bayesian parameter estimation. PGSTAT\footnote{\url{https://heasarc.gsfc.nasa.gov/xanadu/xspec/}} \citep{Arnaud1996ASPC} is utilized as the statistical metric to evaluate the likelihood between the data and model. 
The observational data consist of time-resolved spectra obtained in Section \ref{sec:DATA}. Each burst slice is assigned an observed time $t_{\rm obs,i} ~({\rm i}=1,2,\cdots)$, which corresponds to the midpoint time of the slice, i.e., $t_{\rm obs,i}=(t_{\rm i-1}+t_{\rm i})/2$, where $t_{\rm i-1}$ and $t_{\rm i}$ are the boundary times of slice ${\rm i}$ as tabulated in Table \ref{table:Detectors}. We note that all of these times are referenced to the {\it Fermi}/GBM trigger times.

\begin{figure*}
\centering
\includegraphics[width=1\linewidth]{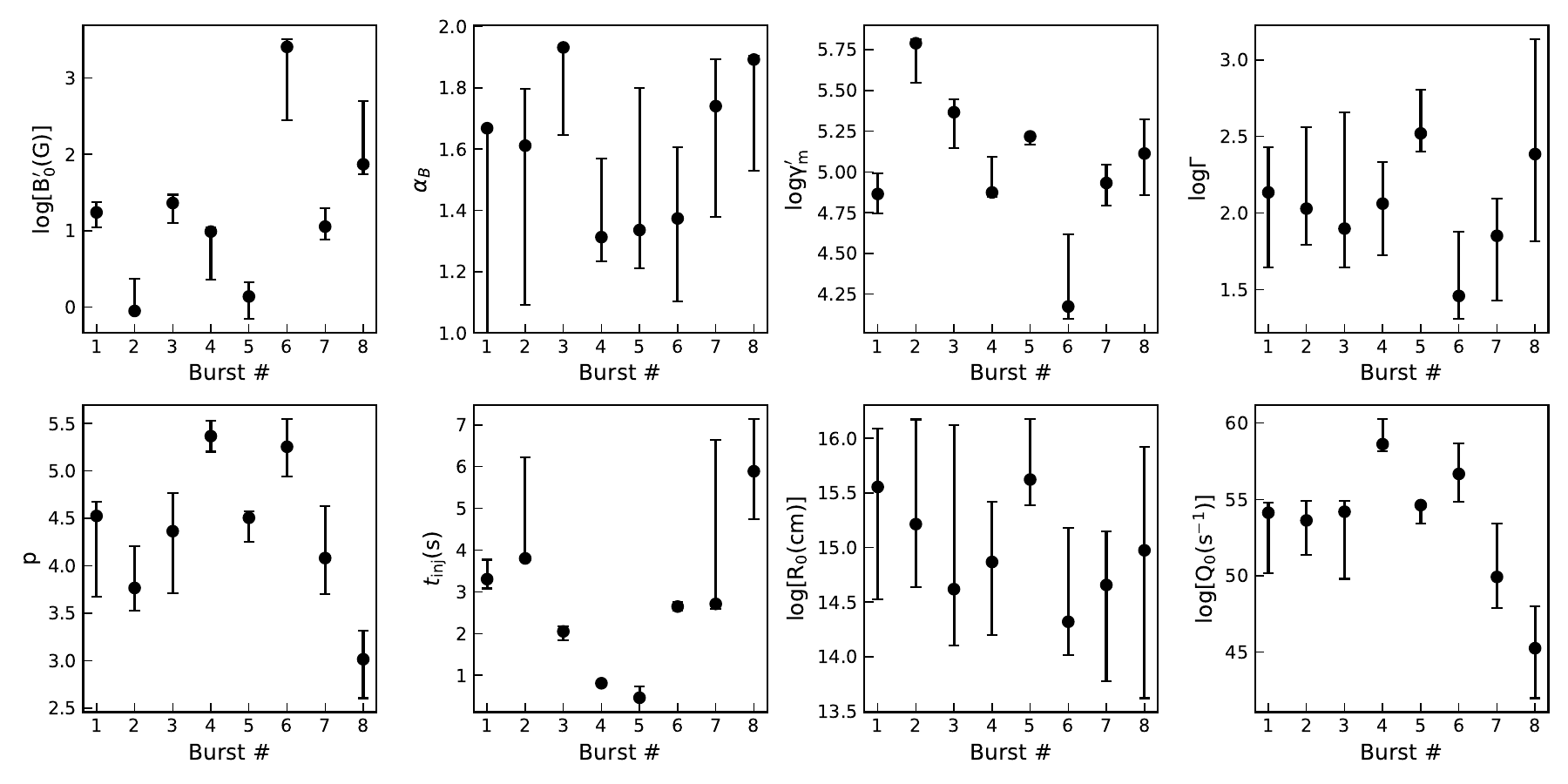}
\caption{Distributions of the best-fit parameters. The burst numbers, 1--8, in the x-axis denote GRB 120326A, GRB 131011A, GRB 140606B, GRB 150514A, GRB 151027A, GRB 170607A, GRB 190829A, and GRB 210204A, respectively. The error bars represent the $1\sigma$ uncertainties. 
\label{fig:parameter_distribution}}
\end{figure*}

For each pulse in our sample, assuming the electron injection starts at the beginning of the 1st time slice, i.e., $t_{\rm s}=t_{0}$, we can employ a single set of free parameters $\mathcal{P}$ to compute a series of theoretical synchrotron spectra using Equation \eqref{eq:Fv_short}. Subsequently, by convolving these spectra with the instrumental response matrix, we can perform a simultaneous fit to the observed time-resolved spectra of the complete set of slices. Therefore, our model is constrained by both the temporal and spectral characteristics of the pulse. By setting appropriate prior bounds for $\mathcal{P}$, as listed in Appendix \ref{sec:C}, we achieve successful fits as listed in Table \ref{table:fit} alongside the best-fit parameters and the goodness of the fits. Figure \ref{fig:parameter_distribution} illustrates the distributions of the best-fit parameters.

Utilizing the best-fit parameters, the model-predicted spectra, light curves, $E_{\rm p}$ evolution, and electron distributions are presented in Figure \ref{fig:120326A} for GRB 120326A and in Appendix \ref{sec:C} for other bursts.
The figures consist of five panels and Figure \ref{fig:120326A} shall be taken as an example to provide a detailed description. In Figure \ref{fig:120326A}a, the observed photon count spectra (data points with error bars) and the model-predicted response-convolved photon count spectra (solid curves) of all slices from all selected detectors are compared. The model-predicted response-convolved light curves (colored curves) of different energy bands from the detector that have the highest S/N are shown in Figure \ref{fig:120326A}b. The observed light curves (gray curves) and the cease time of electron injection (black dashed line) are also plotted in Figure \ref{fig:120326A}b. For the purpose of direct comparison, the light curves in the same energy bands from both the model and observation are scaled and shifted identically. 

In Figure \ref{fig:120326A}c, we compare the $E_{\rm p}$ values from the results of the CPL fits (blue points with error bars) and the synchrotron model fits (orange curve) of the observed spectra in all times slices. The evolution of electron distributions as a function of the source-frame time $2\Gamma ^2 (t-t_0)/(1+z)$ calculated using the best-fit parameters is displayed in Figure \ref{fig:120326A}d. A more detailed evolution process is depicted by the gray curves. Figure \ref{fig:120326A}e presents the evolution of the energy spectra as a function of the observer-frame time calculated using the best-fit parameters. The 1$\sigma$ regions of the posterior uncertainties are presented in shaded areas. The corresponding posterior corner plots of the fitting parameters are displayed in Appendix \ref{sec:C}.

\begin{figure*}
\includegraphics[width=1\linewidth]{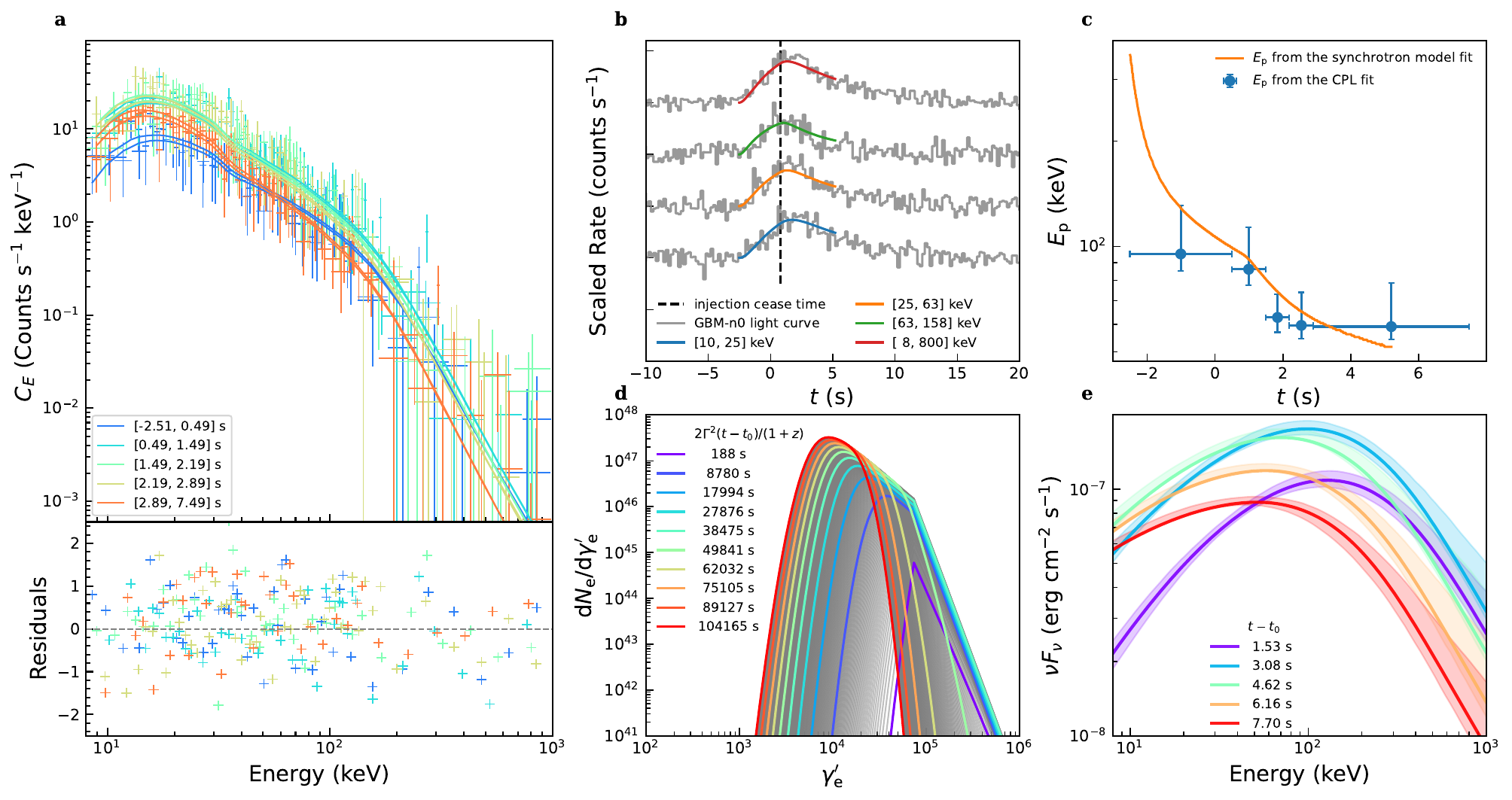}
\caption{The fitting results of the synchrotron model for GRB 120326A. \textbf{a}, The observed (data points with error bars) and modeled (solid curves) photon count spectra. Different curves with the same colors represent the results convolving the response matrices of different detectors within the same time slices. \textbf{b}, The scaled light curves in different energy bands from observation (gray curves) and the synchrotron model fitting results (colored curves). The black dashed line represents the cease time of electron injection. \textbf{c}, The evolution of $E_{\rm p}$ from the CPL fitting results (blue points with error bars) and the synchrotron model fitting results (orange curve). \textbf{d}, The evolution of electron distribution as a function of the source-frame time. The gray curves show more detailed evolution processes. \textbf{e}, The evolution of the energy spectrum ($\nu F_{\nu}$) as a function of the observer-frame time. The error bars and regions represent the $1\sigma$ uncertainties. The complete set of figures (eight images) is available in the online journal.
\label{fig:120326A}}
\end{figure*}

\section{Discussions} \label{sec:Discussion}

\subsection{Electron Distribution} \label{sec:DiscussionElectron}

The observed photons at a specific observer-frame time were emitted from the electrons at distinct source-frame times and locations on the EATS. Consequently, considering the curvature effect from high latitudes on the EATS, these observed photons are not solely produced by the electron distribution at a single source-frame time. This rationale justifies our decision to illustrate the evolution of the electron distribution in Figure \ref{fig:120326A} 
using the source-frame time rather than the observer-frame time. In the initial stages of the calculations, corresponding to the earliest times, the electron distributions usually exhibit a sharp profile, which to some extent reflects the initial morphology of electron injection.

Figure \ref{fig:120326A} 
indicates that the cooling of electrons is rapid prior to the cessation of electron injection, while it becomes more pronounced after that. In particular, the fitting results of GRB 170607A present an extremely fast cooling rate. 
This is because, compared with other GRBs' fitting results in Table \ref{table:fit}, its higher initial magnetic field strength $B^{\prime}_{\rm 0} = 10^{3.41}$ G and lower magnetic field decay index $\alpha_B=1.37$ lead to efficient synchrotron cooling, according to Equation \eqref{eq:cooling}. Moreover, its relatively small initial radius $R_{\rm 0}=10^{14.32} \ {\rm cm}$ enhances the initial adiabatic cooling efficiency, while the comparatively low bulk Lorentz factor, $\Gamma=10^{1.46}$, sustains a high cooling rate throughout the pulse duration.

In contrast, the pulse of GRB 190829A exhibits considerable inefficiency in the cooling process. This can be attributed to its weak magnetic field and relatively short duration, both of which impose constraints on the available time for electrons to undergo significant cooling. As a consequence, the electron distribution retains its shape to a large extent throughout the pulse.

\subsection{{Synchrotron Cooling vs. Adiabatic Cooling}}

Electron cooling in this paper involves both synchrotron and adiabatic cooling. The synchrotron cooling rate is generally much larger than the adiabatic cooling rate at an early co-moving time (or a small radius). However, the former decays faster compared to the latter. This is attributed to the fact that $\left({\rm d} \gamma^{\prime}_{\rm e}/{\rm d} R\right)_{\rm syn} \propto {B^{\prime}}^2 \propto R^{-2\alpha_B}$, while $\left({\rm d} \gamma^{\prime}_{\rm e}/{\rm d} R\right)_{\rm adi} \propto R^{-1}$. It is important to note that in this work, $\alpha_B > 1/2$ for all the bursts in our sample. Thus, we can determine the transition radius $R_{\rm adi}$ above which adiabatic cooling becomes dominant by comparing the synchrotron and adiabatic cooling rates:

\begin{equation}
\label{eq:R_adi}
R_{\rm adi} = \left(\frac{\sigma_{\rm T}\gamma^{\prime}_{\rm e} B_0^{\prime 2} R_0^{2\alpha_B}}{4\pi m_{\rm e} c^2 \beta \Gamma} \right)^{1/(2\alpha_B-1)}.
\end{equation}

Such a critical radius is dependent on the electron energy, $\gamma^{\prime}_{\rm e}$. We can set $\gamma'_{\rm e}=\gamma^{\prime}_{\rm m}$ to calculate $R_{\rm adi}$ since the electron energy concentrates on $\gamma^{\prime}_{\rm m}$, which is the minimal Lorentz factor in the electron distribution defined in Equation \eqref{eq:injection}.

In terms of an observer's perspective, it is essential to determine the portion of the light curve primarily influenced by either adiabatic or synchrotron cooling. The observed flux for a given observed time is from the EATS that covers all radii, including both the regions dominated by the synchrotron cooling and those dominated by the adiabatic cooling. In order to determine when the adiabatic cooling dominates in light curves, we need to find the time after which the contribution from the adiabatic cooling electrons dominates over that from the synchrotron cooling electrons. Therefore, we introduce a novel ratio defined as the percentage of the flux ($F_{\rm adi}$) from the adiabatic loss-dominating region of the EATS ($R>R_{\rm adi}$) to the flux ($F_{\rm tot}$) from the total EATS ($R>R_0$), i.e.,
\begin{equation} \label{eq:frac}
 f=\frac{F_{\rm adi}(R_{\rm adi})}{F_{\rm tot}}.
\end{equation}
Once this ratio exceeds 0.5 at a specific observer-frame time $t_{\rm tr}$, it is reasonable to regard this time as the transition time beyond which the adiabatic loss dominates over the synchrotron loss. 

Utilizing the best-fit parameters, we first calculate the radius $R_{\rm adi}$ for each burst. We identify three distinct cases based on the relationship between $R_{\rm adi}$, $R_0$, and $R_{\rm max}$. Here $R_{\rm max}$ is the largest radius in the EATS for the middle time of the maximum time slice. 
We find that for GRB 140606B, GRB 150514A, GRB 190829A, and GRB 210204A, the relation $R_0<R_{\rm adi}<R_{\rm max}$ holds. Using Equation \eqref{eq:frac}, we calculate the ratio $f$ in 1--10000 keV energy range to determine the transition time $t_{\rm tr}$, which marks the time point where the adiabatic cooling dominates over the synchrotron cooling. For GRB 150514A, GRB 190829A, and GRB 210204A, the transition times are $2.21$, $1.16$, and $37.34~{\rm s}$, respectively. In the case of GRB 140606B, the value of $f$ remains below 0.5 during the entire light curve, suggesting that synchrotron cooling remains dominant throughout the evolution of the light curve. Similarly, for GRB 120326A and GRB 170607A, due to $R_{\rm adi}>R_{\rm max}$, the transition times are larger than their duration, suggesting that the synchrotron cooling is dominant throughout their light curves as well. Therefore, the ratio $f\sim0$ for the two bursts.

For GRB 131011A and GRB 151027A, we find that the transition radius is exceptionally small with $R_{\rm adi}<R_0$. This suggests that the initial magnetic field strength or synchrotron cooling rate of the two bursts is comparatively low. As a result, the adiabatic losses dominate their light curves from the beginning of the two bursts, leading to $f\sim1$. This implies that the electrons in the two bursts radiate inefficiently, raising concerns about a potential energy budget problem, as low radiation efficiency necessitates a high total energy budget. We calculate the total injected electron energy using the best-fit parameters for the two bursts, resulting in $1.11\times10^{54}$ and $6.38\times10^{52}~{\rm erg}$. These values fall within an acceptable range.

\begin{figure*}
\centering
\includegraphics[width=1\linewidth]{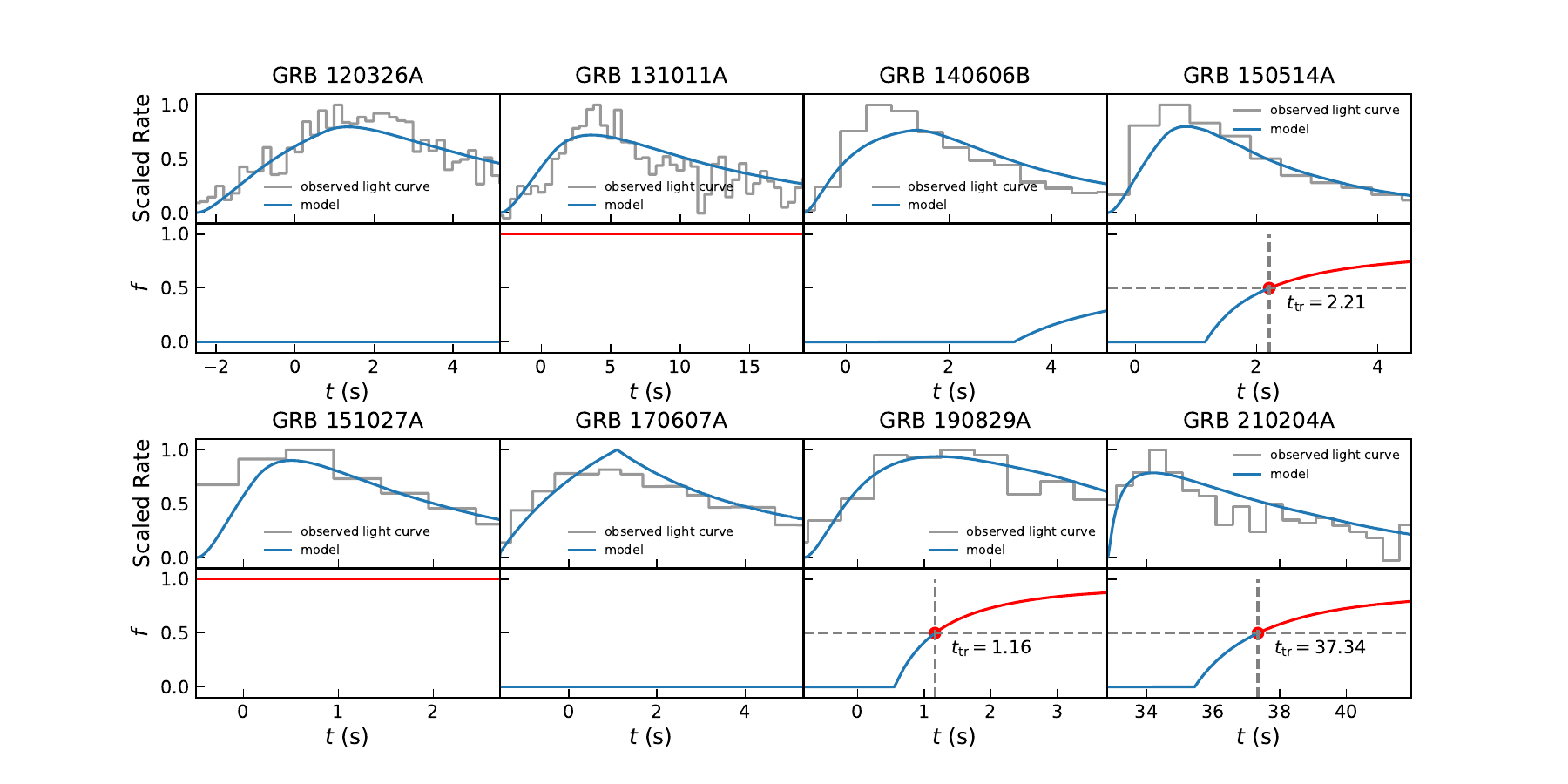}
\caption{The scaled light curves and the evolution of the ratios of the flux from the adiabatic loss-dominating regions of the EATS to the flux from the total EATS. In the lower panels, the times when the synchrotron losses dominate are depicted in blue, while the times when the adiabatic losses dominate are depicted in red. The horizontal gray dashed lines display where the ratio is equal to 0.5.
\label{fig:adiabatic_ratio}}
\end{figure*}

In Figure \ref{fig:adiabatic_ratio}, we depict the evolution of the ratio $f$ and the light curves for all the bursts. The upper panels of the figure display the scaled model-predicted light curves and observed light curves, which are the same as the light curves with the widest energy bands on the top of Figure \ref{fig:120326A}b. The lower panels show the ratio evolution. The times when the synchrotron losses dominate (ratio $<$ 0.5) are depicted in blue, while the times when the adiabatic losses dominate (ratio $>$ 0.5) are displayed in red. We can find that the light curves after the adiabatic cooling becomes dominant have no significant distinction compared with those before the time. This can be attributed to the effect of the EATS, which tends to smear the difference between the two cooling components. It is also possible that the curvature effect plays a significant role in shaping the light curves. We will discuss this problem further in subsection \ref{sec:DiscussionCurvature}.

\subsection{$E_{\rm p}$ Evolution} \label{sec:DiscussionEp}

The energy spectral peak of GRBs in the synchrotron model is 
\bea
E_{\rm p}\propto \Gamma {\gamma'_{\rm m}}^2 B'.
\eea
In our model, the bulk Lorentz factor $\Gamma$ is held as a constant parameter, signifying a scenario where the shell expands at a constant speed. The electron injection in the shell is uniform in both space and time and $\gamma'_{\rm m}$ is set to be a constant. The magnetic field strength $B^{\prime}$ decreases monotonically as the shell expands. Hence, our model is only capable of producing GRBs with the hard-to-soft $E_{\rm p}$ evolution pattern.
This constitutes the basis for our selection of GRBs that exhibit this specific pattern, as detailed in the spectral screening process in Section \ref{sec:DATA}. 
 
A small fraction of bursts also exhibit the behavior of ``$E_{\rm p}$ tracking the flux" rather than the hard-to-soft pattern \citep[e.g.,][]{2012ApJ...756..112L}, implying that additional effects should be taken into account. These effects can include the acceleration of the emitting region \citep{2015ApJ...808...33U,2016ApJ...825...97U,2018ApJ...869..100U,2021A&A...656A.134G} and the synchrotron self-Compton process \citep{2014ApJ...780...12Z, 2018ApJS..234....3G,2021A&A...656A.134G}. Our findings demonstrate that at least for some bursts, synchrotron radiation alone can serve as a dominant factor and consistently reproduce the observed $E_{\rm p}$ evolution. Further, employing the best-fit parameters, we calculate the spectral lags for the sample pulses, and all results exhibit positive lags. This is aligned with the prediction made by \cite{2018ApJ...869..100U} that for the hard-to-soft $E_{\rm p}$ pattern, only the positive type of spectral lags is possible.

\subsection{Temporal Properties} \label{sec:DiscussionTemporal}
In Figures \ref{fig:120326A}b 
, the majority of the cease times of electron injection coincide with the peak times of the light curves. 
This finding corroborates the anticipated conclusion that the rising segment of the FRED shape is the result of electron injection and accumulation, while the decaying segment is attributable to the cooling of electrons and the curvature effect. 

However, an interesting exception is observed for the first episode of GRB 190829A, as illustrated in Figure \ref{fig:120326A}b. In this case, the cessation of electron injection occurs considerably later than the temporal peak time. This discrepancy is due to the rapid decay of the magnetic field, leading to inefficient synchrotron radiation. As a result, even if there is a continuing injection and accumulation of electrons, the photon emission experiences a decline during the later stages before the injection eventually ceases. The fitting results presented in Table \ref{table:fit} confirm that the pulse of GRB 190829A possesses distinct characteristics, including a lower initial magnetic field, a higher magnetic field decay index, and a lower initial radius, when compared with the other sample pulses. These features collectively contribute to the inefficiency of synchrotron radiation at later times. Furthermore, as discussed in Section \ref{sec:DiscussionElectron}, the electron distributions of GRB 190829A at late times indeed display inefficient cooling behavior.

\subsection{Curvature Effect} \label{sec:DiscussionCurvature}
\begin{figure}[h!]
\centering
\includegraphics[width=0.95\linewidth]{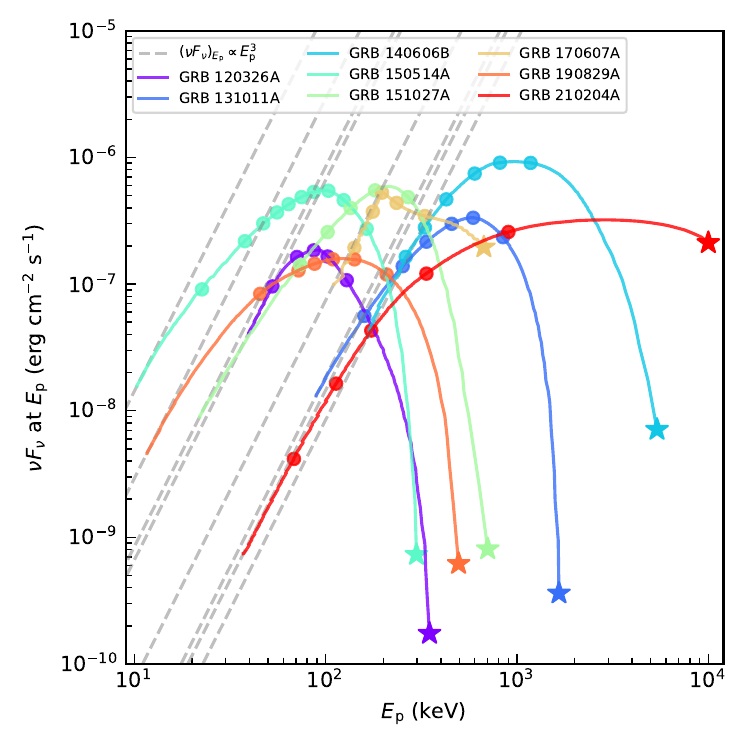}
\caption{$(\nu F_{\nu})_{E_{\rm p}}$ and $E_\mathrm{p}$ correlation diagram calculated using the best-fit parameters. The pentagrams indicate the initial values of $(\nu F_{\nu})_{E_{\rm p}}$ and $E_\mathrm{p}$ calculated at $t_0$ and the circles indicate the values calculated at the midpoint times of the time intervals. The correlations exhibit a time evolution, which ultimately converges to the anticipated correlation of high-latitude emission with $(\nu F_{\nu})_{E_{\rm p}} \propto {E_{\rm p}}^3$ as presented in gray dashed lines. 
\label{fig:vFv_Ep}}
\end{figure}

The correlation between the flux at the peak energy and the peak energy, characterized as $F_{\nu,E_{\rm p}} \propto E_{\rm p}^2$, has been predicted for high-latitude emission \citep{2004ApJ...614..284D,2021ApJS..253...43L,2022arXiv221207094U}. Such a correlation aligns with the results of \cite{2023ApJ...949..110T}, where the exponent $\delta$ in the relation $F_{\nu,E_{\rm p}} \propto E_{\rm p}^\delta$ is found to follow a Gaussian distribution with a median value of 1.99 and a width of 0.34 for the 18 broad pulses in their sample. Consequently, a related correlation, expressed as $(\nu F_{\nu})_{E_{\rm p}} \propto E_{\rm p}^3$, should also exist. To investigate when and how the high-latitude emission comes into play within our model, we present the $(\nu F_{\nu})_{E_{\rm p}}$ and $E_{\rm p}$ correlations with the best-fit parameters in Figure \ref{fig:vFv_Ep}. The pentagrams indicate the initial values calculated at $t_0$ and the circles indicate the values calculated at the midpoint times of the time intervals. The gray dashed lines illustrate the $(\nu F_{\nu})_{E_{\rm p}} \propto E_{\rm p}^3$ correlation. Notably, in Figure \ref{fig:vFv_Ep}, all the best-fit results converge to this correlation at late times when high-latitude emissions dominate. 

It is interesting that this correlation is only satisfied at the late time, which is much later than the temporal peak time and even later than the last time interval listed in a few cases: GRB 150514A, GRB 151027A, and GRB 190829A. Therefore, for these cases, the data used to perform the fit may not be sufficiently late to probe the high-latitude emission, and this correlation is to some extent only a prediction of the model.

We note that the light curves for both adiabatic cooling-dominated and synchrotron cooling-dominated bursts are subject to the curvature effect. This implies that the contribution to the flux from the adjacent area of the LOS for these bursts is always smaller than that from the high-latitude region. However, it is possible that, under certain parameter conditions, adiabatic cooling can prevail over the curvature effect in some bursts. Additionally, if the jet effect is considered, the contribution from the high-latitude region will decline to zero at a certain time due to the arrival of the jet edge. The light curve after this time will be solely determined by adiabatic-cooling electrons. Such cases will be further studied in our future research.

\section{Summary} \label{sec:Conclusions}
In this study, we have presented a novel and comprehensive approach for theoretically understanding the temporal and spectral evolution of GRBs within the framework of the synchrotron model. Our method involves fitting the time-resolved spectra of GRBs using a unified set of parameters derived from the synchrotron model. By applying this approach to a sample of eight GRBs characterized by FRED shapes, we have demonstrated the exceptional performance of our model in fitting the observed time-resolved spectra. Notably, the evolution of the spectral peak $E_{\rm p}$ and the light curve, as predicted based on the best-fit parameters, exhibits remarkable agreement with the actual observational data. Our results provide strong evidence that the intricate spectral and temporal evolution in GRBs is predominantly governed by the dynamic evolution of key physical quantities, such as magnetic field strength and electron distribution, within the emission region as a consequence of its expansion. Our results also lend support to the notion that GRB emission occurs at a typical radius of approximately $10^{15}$ cm, aligning with the magnetic dissipation model and further affirming the synchrotron emission mechanism as the underlying radiation process.

\begin{acknowledgments}
We acknowledge the support by the National Key Research and Development Programs of China (2022YFF0711404, 2022SKA0130102), the National SKA Program of China (2022SKA0130100), the National Natural Science Foundation of China (grant Nos. 11833003, U2038105, U1831135, 12121003, 12393811), the science research grants from the China Manned Space Project with NO. CMS-CSST-2021-B11, the Fundamental Research Funds for the Central Universities, the Program for Innovative Talents and Entrepreneur in Jiangsu, the Chinese Academy of Sciences, grant No. XDB23040400, and the Postgraduate Research \& Practice Innovation Program of Jiangsu Province (KYCX23\_0117). This work was performed on an HPC server equipped
with two Intel Xeon Gold 6248 modules at Nanjing
University. We acknowledge IT support from the computer
lab of the School of Astronomy and Space Science at Nanjing University. 
\end{acknowledgments}

\appendix
\counterwithin{figure}{section}
\counterwithin{table}{section}
\counterwithin{equation}{section}

\section{The Constrained Interpolation Profile method}
\counterwithin{figure}{section}
\counterwithin{table}{section}
\counterwithin{equation}{section}
\label{sec:CIP}

To solve Equation \eqref{eq:electron}, we perform the constrained
interpolation profile (CIP) method \citep{2001JCoPh.169..556Y,2018ApJS..234....3G}. First, the equation can be transformed into
\begin{equation} \label{eq:electron_CIP}
\frac{\partial}{\partial R} \left(\frac{{\rm d}N_{\rm e}^{\prime}}{{\rm d}x} \right) +
\frac{\partial}{\partial x} \left[ \frac{{\rm d} x}{{\rm d} R} \left(\frac{{\rm d}N_{\rm e}^{\prime}}{{\rm d} x} \right) \right]=Q^{\prime}\left( x \right) \gamma_{\rm e}^{\prime} \ln{10},
\end{equation}
where $x=\log_{10} \gamma_{\rm e}^{\prime}$. Then the equation can be abstracted as 
\begin{equation} \label{eq:electron_abstracted}
\frac{\partial f}{\partial R} +
\frac{\partial \left(u f\right)}{\partial x} = g
\end{equation}
where $f={\rm d}N_{\rm e}^{\prime} /{\rm d}x $ is the quantity that we need to solve, $u={\rm d} x / {\rm d} R=-\sigma_{\rm T} {B^{\prime}}^2 \gamma_{\rm e}^{\prime} /(6 \ln{10} \pi m_{\rm e} c^2 \beta \Gamma) - 2/(3\ln{10} R)$ according to Equation \eqref{eq:cooling} and $g=\ln{10} Q^{\prime}\left( x \right) \gamma_{\rm e}^{\prime} $. The solution procedure can be separated into two fractional steps, the convection phase and the non-convection phase,
\begin{equation} \label{eq:convection}
{\rm convection}
\left\{ 
\begin{aligned}
\frac{\partial f}{\partial R} +
u \frac{\partial f}{\partial x} = 0, \\
\frac{\partial \hat{f}}{\partial R} +
u \frac{\partial \hat{f}}{\partial x} = 0, 
\end{aligned}
\right. 
\end{equation}

\begin{equation} \label{eq:nonconvection}
{\rm non-convection}
\left\{ 
\begin{aligned}
&\frac{\partial f}{\partial R} =
g - f \frac{\partial u}{\partial x}, \\
&\frac{\partial \hat{f}}{\partial R} =
\frac{\partial g}{\partial x} - f \frac{\partial^2 u}{\partial x^2} - 2\hat{f} \frac{\partial u}{\partial x}, 
\end{aligned}
\right. 
\end{equation}
where $\hat{f}=\partial f/ \partial x$, $\partial u/ \partial x=-\sigma_{\rm T} {B^{\prime}}^2 \gamma_{\rm e}^{\prime} /(6\pi m_{\rm e} c^2 \beta \Gamma)$, $\partial^2 u/\partial x^2 = - \ln{10} \sigma_{\rm T} {B^{\prime}}^2 \gamma_{\rm e}^{\prime} /(6\pi m_{\rm e} c^2 \beta \Gamma) = \ln{10} \partial u/ \partial x $, and $\partial g / \partial x = (-p+1)(\ln{10})^2 Q^{\prime}\left( x \right) \gamma_{\rm e}^{\prime} $. Note that we calculate the partial components above in the exact expressions, but one can also replace them with some difference schemes \citep[e.g., the centered finite difference scheme;][]{2018ApJS..234....3G}. We use the conventions that the superscripts ${\rm \cdots i,i+1 \cdots }$ represent the computing grid numbers for $R$ and the subscripts ${\rm \cdots j,j+1 \cdots} $ represent the computing grid numbers for $x$. Using the non-convection phase in Equation \eqref{eq:nonconvection}, we can get intermediate solutions $f_{\rm j}^{*}$ and $\hat{f}_{\rm j}^{*}$: 
\begin{equation} \label{eq:nonconvection_intermediate}
\left\{ 
\begin{aligned}
&f_{\rm j}^{*} = f_{\rm j}^{\rm i}
+ {\left( g - f \frac{\partial u}{\partial x} \right)}_{\rm j}^{\rm i} \Delta R, \\
&\hat{f}_{\rm j}^{*} = {\hat{f}_{\rm j}}^{\rm i}
+{ \left( \frac{\partial g}{\partial x} - f \frac{\partial^2 u}{\partial x^2} - 2\hat{f} \frac{\partial u}{\partial x} \right) }_{\rm j}^{\rm i} \Delta R, 
\end{aligned}
\right. 
\end{equation}

Assume that the intermediate solutions $f^{*}$ and $\hat{f}^{*}$ of two adjacent points ${\rm j}$ and ${\rm j+1}$ on the $x$-axis satisfy cubic polynomial approximations:
\begin{equation} \label{eq:cubic_polynomial}
\left\{ 
\begin{aligned}
&f_{\rm j}^{*} = A_{\rm j}^{*}{x_{\rm j}}^3 + B_{\rm j}^{*}{x_{\rm j}}^2 + C_{\rm j}^{*}x_{\rm j} + D_{\rm j}^{*} \\
&f_{\rm j+1}^{*} = A_{\rm j}^{*}{(x_{\rm j}+\Delta x)}^3 + B_{\rm j}^{*}{(x_{\rm j}+\Delta x)}^2 + C_{\rm j}^{*}(x_{\rm j}+\Delta x) + D_{\rm j}^{*} \\
&\hat{f}_{\rm j}^{*} = 3A_{\rm j}^{*}{x_{\rm j}}^2 + 2B_{\rm j}^{*}x_{\rm j} + C_{\rm j}^{*} \\
&\hat{f}_{\rm j+1}^{*} = 3A_{\rm j}^{*}{(x_{\rm j}+\Delta x)}^2 + 2B_{\rm j}^{*}(x_{\rm j}+\Delta x) + C_{\rm j}^{*}.
\end{aligned}
\right. 
\end{equation}
Then, considering the convection phase, the final solutions should be just the x-axis shifts of the non-convection solutions (intermediate solutions) with shift distance $u \Delta R$:
\begin{equation} \label{eq:convection_shift_final}
\left\{ 
\begin{aligned}
f_{\rm j}^{\rm i+1} =& A_{\rm j}^{*}{(x_{\rm j}-u \Delta R)}^3 + B_{\rm j}^{*}{(x_{\rm j}-u \Delta R)}^2 \\
 &+ C_{\rm j}^{*}(x_{\rm j}-u \Delta R) + D_{\rm j}^{*}\\
 =&-a_{\rm j}^{*}u^3\Delta {R}^3+b_{\rm j}^{*}u^2\Delta {R}^2-\hat{f}_{\rm j}^{*}u\Delta R+f_{\rm j}^{*} \\
\hat{f}_{\rm j}^{\rm i+1} =& 3A_{\rm j}^{*}{(x_{\rm j}-u \Delta R)}^2 + 2B_{\rm j}^{*}(x_{\rm j}-u \Delta R) + C_{\rm j}^{*} \\
 =&3a_{\rm j}^{*}u^2\Delta {R}^2-2b_{\rm j}^{*}u\Delta R+\hat{f}_{\rm j}^{*}.
\end{aligned}
\right. 
\end{equation}
where 
\begin{equation} \label{eq:cubic_coefficients_intermediate}
\left\{ 
\begin{aligned}
&a_{\rm j}^{*} \equiv \frac{\hat{f}_{\rm j}^{*}+\hat{f}_{\rm j+1}^{*}} {\Delta x^2} -2\frac{f_{\rm j+1}^{*}-f_{\rm j}^{*}}{\Delta x^3} \\
&b_{\rm j}^{*}\equiv 3\frac{f_{\rm j+1}^{*}-f_{\rm j}^{*}}{\Delta x^2} - \frac{2\hat{f}_{\rm j}^{*}+\hat{f}_{\rm j+1}^{*}}{\Delta x},
\end{aligned}
\right. 
\end{equation}

The convection phase requires the time step $\Delta R$ in Equation \eqref{eq:convection_shift_final} to satisfy the Courant condition:
\begin{equation} \label{eq:Courant}
\Delta R \leq \left( \frac{\Delta \gamma_{\rm e}^{\prime}}{|{\rm d}\gamma_{\rm e}^{\prime}/{\rm d}R|}\right) _{\rm min}.
\end{equation}

Setting a zero initial condition $f^{\rm 0}=\hat{f}^{\rm 0}=0$, an electron injection source in Equation \eqref{eq:injection}, an electron cooling rate in Equation \eqref{eq:cooling}, and other free parameters in Equation \eqref{eq:Fv_total} and calculating Equations \eqref{eq:nonconvection_intermediate}, \eqref{eq:cubic_coefficients_intermediate}, and \eqref{eq:convection_shift_final}, the electron distribution evolution in each radius step (or each time step) can be obtained. Note that although the CIP method can provide precise numerical solutions, nonphysical oscillation may be encountered below the electrons' lowest energy region. These nonphysical values can be set as zeros manually to avoid the numerical oscillation caused by the explicit difference scheme. In any case, these nonphysical values are too low to change the total electron number. Moreover, they only exist in the lowest energy region, so they have little influence on the cooling processes of
upstream electrons with higher energy.

\section{The fully implicit difference Method} \label{sec:B}
\counterwithin{figure}{section}
\counterwithin{table}{section}
\counterwithin{equation}{section}

The second numerical method that we use to solve Equation \eqref{eq:electron} is proposed in \cite{1999MNRAS.306..551C}. In this method, a fully implicit difference scheme is used, which is a modification of the scheme proposed by \cite{1970JCoPh...6....1C}.
As done in \cite{1999MNRAS.306..551C}, we use an energy grid with
equal logarithmic resolution, where the energy mesh points are defined as
 \bea
\gamma_{\rm{e,j}}^{\prime}=\gamma_{\rm {e,min}}^{\prime}(\gamma_{\rm {e,max}}^{\prime}/\gamma_{\rm {e,min}}^{\prime})^{(\rm{j-1})/(\rm{j}_{\rm{max}}-1)},
 \eea
where ${\rm j}_{\rm {max}}$ is the mesh-point number, and the energy interval
is $\Delta \gamma_{\rm {e,j}}^{\prime}=\gamma_{\rm {e,j+1/2}}^{\prime}-\gamma_{\rm {e,j-1/2}}^{\prime}$. The continuity equation can be written as
 \bea
\f{N_{\rm e,j}^{\prime, \rm i+1}-N_{\rm e,j}^{\prime, \rm i}}{\Delta
R}+\f{F^{\rm i+1}_{\rm j+1/2}-F^{\rm i+1}_{\rm j-1/2}}{\Delta\gamma_{\rm {e,j}}^{\prime}}=Q^{\prime, \rm i}_{\rm j}\\
F^{\rm i+1}_{\rm j\pm1/2}=(C^{\rm i}_{\rm syn,j\pm1/2}+C^{\rm i}_{\rm adi,j\pm1/2})N^{\prime, \rm i+1}_{\rm e,j\pm1/2}.
 \eea
Here $C_{\rm{syn}}+C_{\rm{adi}}={\rm d}\gamma_{\rm e}^{\prime}/{\rm d} R$ is the sum of the synchrotron and adiabatic cooling rates. 
We define $N_{\rm e,j+1/2}^{\prime}\equiv N_{\rm e,j+1}^{\prime}$ and
$N_{\rm e,j-1/2}^{\prime}\equiv N_{\rm e,j}^{\prime}$ and then rewrite the continuity equation as
 \bea \label{discrete_eq}
V_{\rm 3,j}^{\rm i}N_{\rm e,j+1}^{\prime, \rm i+1}+V_{\rm 2,j}^{\rm i}N_{\rm e,j}^{\prime, \rm i+1}+V_{\rm 1,j}^{\rm i}N_{\rm e,j-1}^{\prime, \rm i+1}=r_{\rm j}^{\rm i}, 
 \eea
where $V_{\rm 1,j}^{\rm i}=0$, 
$V_{\rm 2,j}^{\rm i}=1-\Delta R C^{\rm i}_{\rm j-1/2}/\Delta\gamma_{\rm e,j}^{\prime}$, 
$V_{\rm 3,j}^{\rm i}=\Delta R C^{\rm i}_{\rm j+1/2}/\Delta\gamma_{\rm j}^{\prime}$, and $r_{\rm j}^{\rm i}= Q_{\rm j}^{\prime, \rm i}\Delta R+N_{\rm e,j}^{\prime, \rm i}$. A tridiagonal matrix is derived from Equation \eqref{discrete_eq} and solved numerically \citep{1992nrfa.book.....P}. 

\hspace*{\fill}

Following extensive testing, we have found that the codes developed using the two methods presented in Appendices \ref{sec:CIP} and \ref{sec:B} yield essentially consistent results, thereby ensuring the accuracy and reliability of our approaches. The relevant results pertinent to this study are produced by the code using the methodology described in Appendix \ref{sec:B}.

\section{Additional Tables and Figures}
\label{sec:C}
\counterwithin{figure}{section}
\counterwithin{table}{section}
\counterwithin{equation}{section}

The selected detectors and the segmented time slices of the final sample are listed in Table \ref{table:Detectors}.

The prior bounds for the fits are listed in Table \ref{table:fit_bounds}.

\begin{figure}[h!]
\plotone{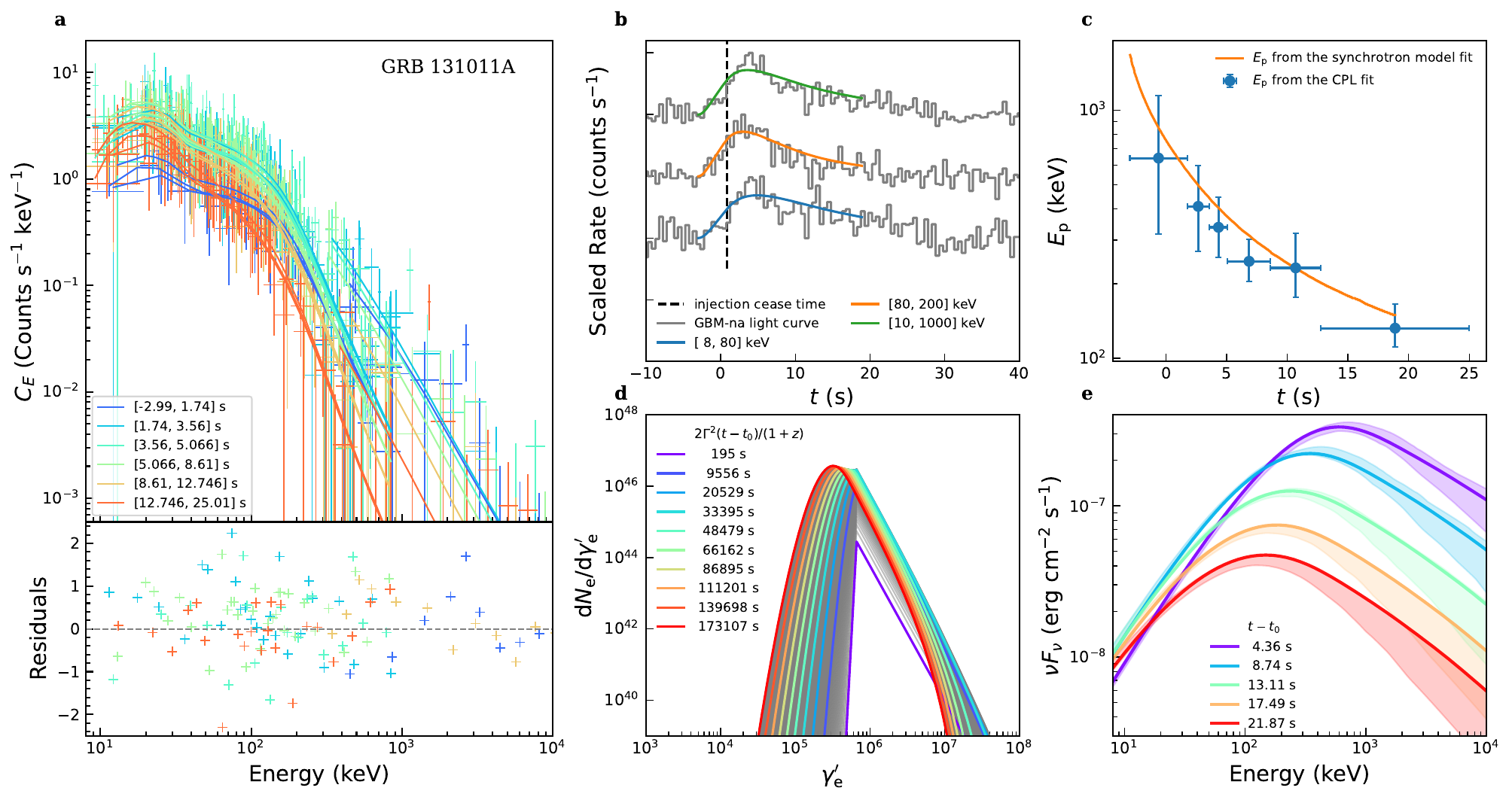}
\caption{Same as Figure \ref{fig:120326A}, but for GRB 131011A.
\label{fig:131011A}}
\end{figure}

\begin{figure}[h!]
\plotone{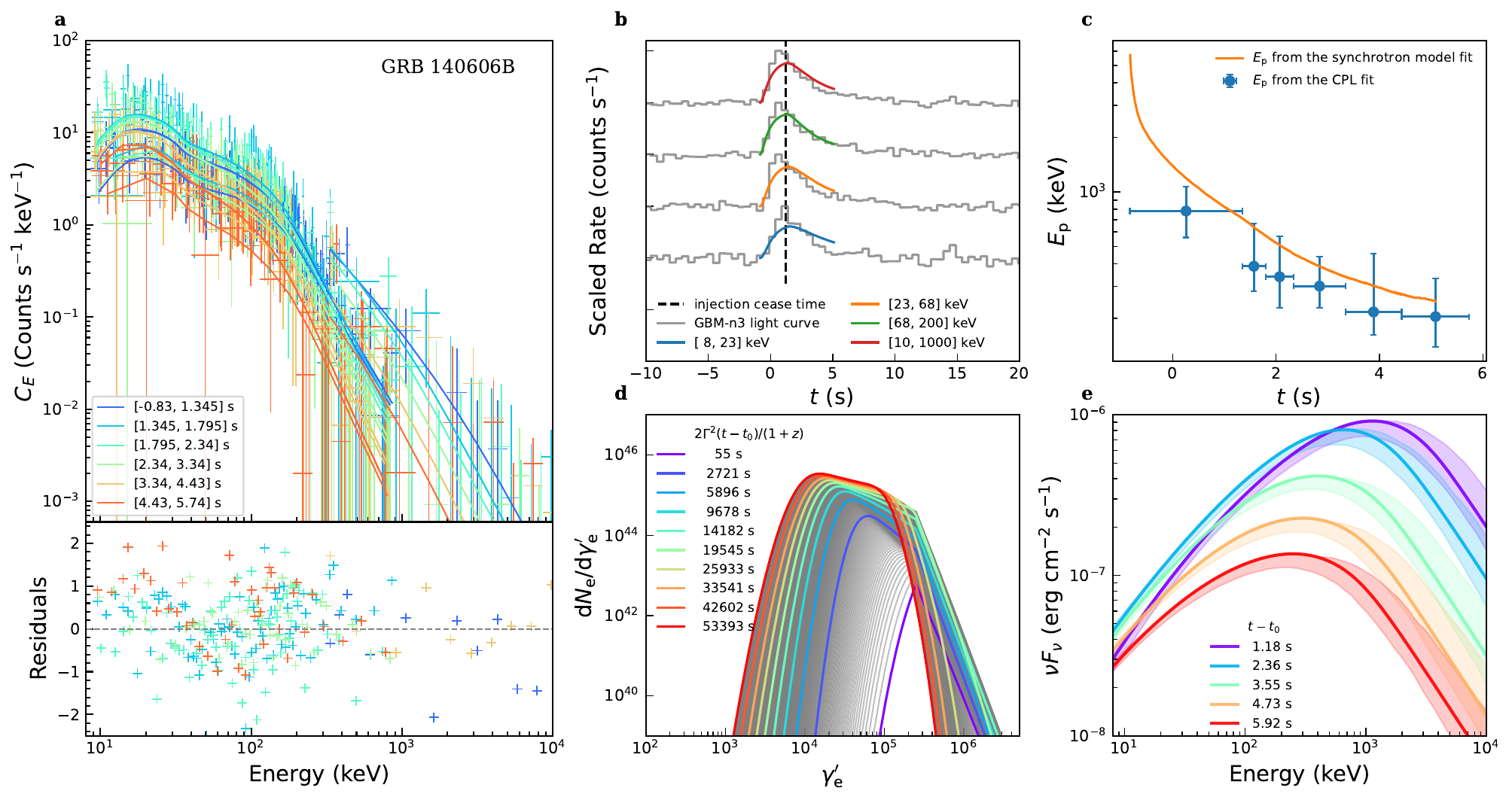}
\caption{Same as Figure \ref{fig:120326A}, but for GRB 140606B.
\label{fig:140606B}}
\end{figure}

\begin{figure}[h!]
\plotone{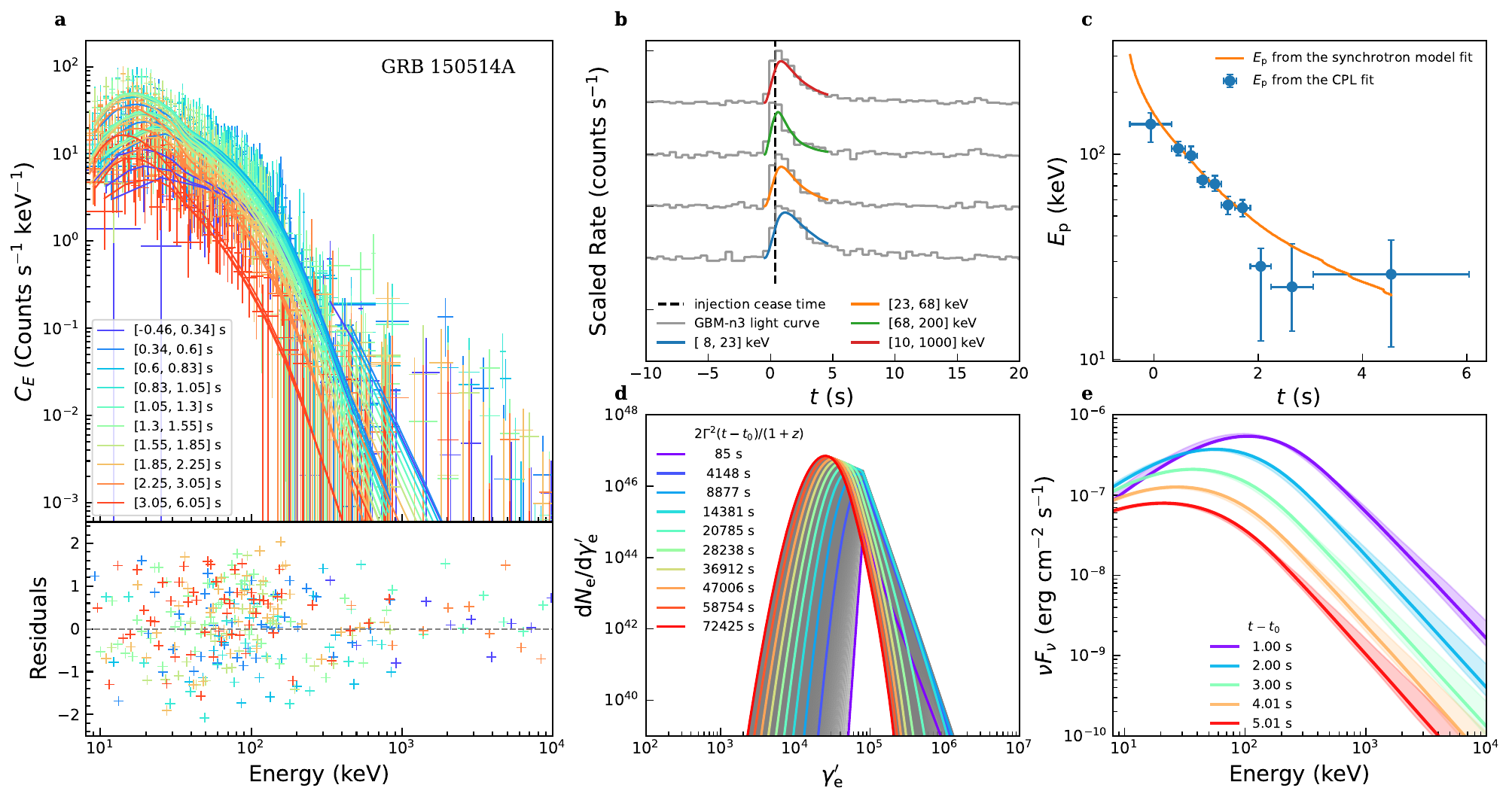}
\caption{Same as Figure \ref{fig:120326A}, but for GRB 150514A.
\label{fig:150514A}}
\end{figure}

\begin{figure}[h!]
\plotone{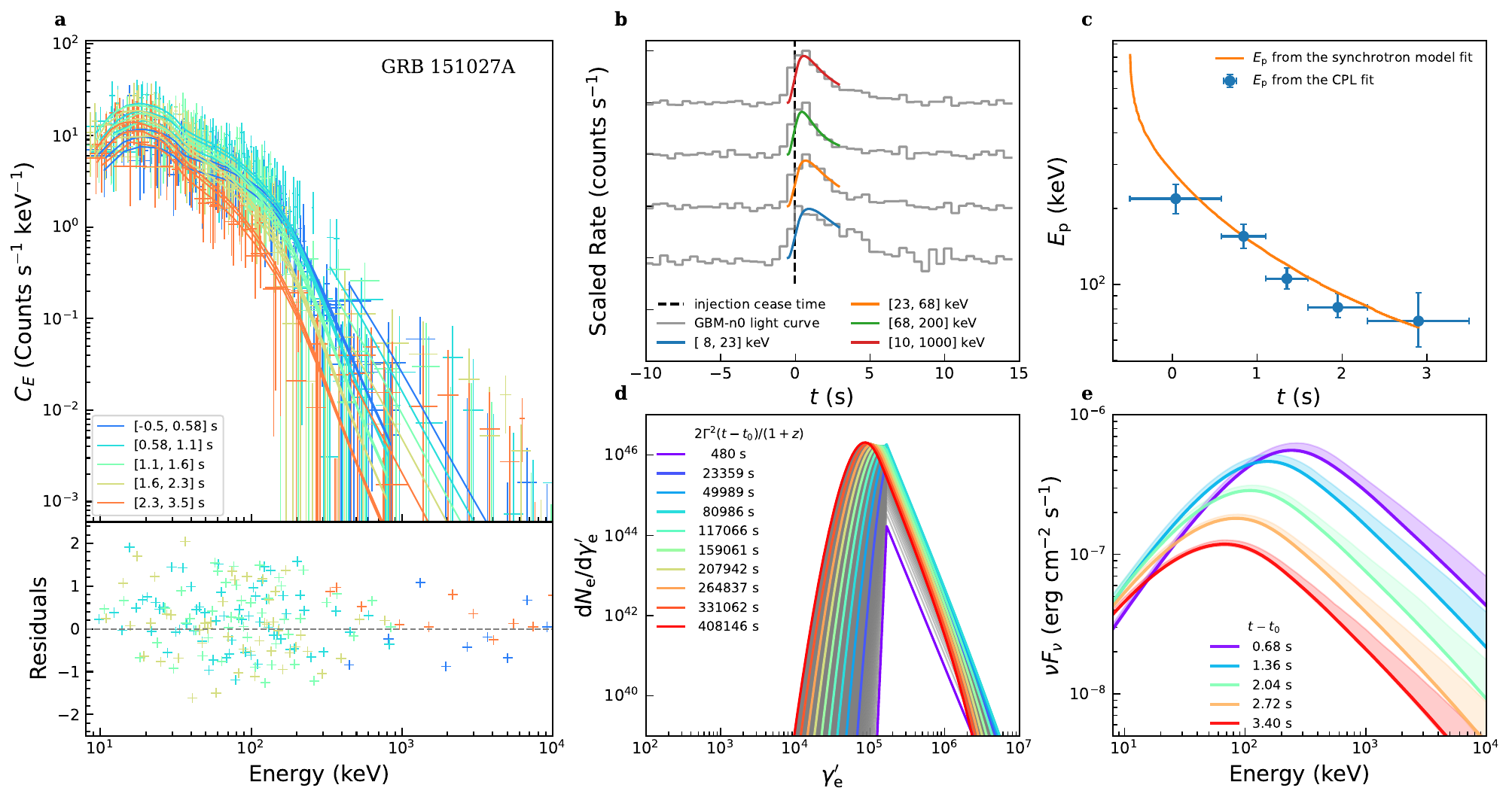}
\caption{Same as Figure \ref{fig:120326A}, but for GRB 151027A.
\label{fig:151027A}}
\end{figure}

\begin{figure}[h!]
\plotone{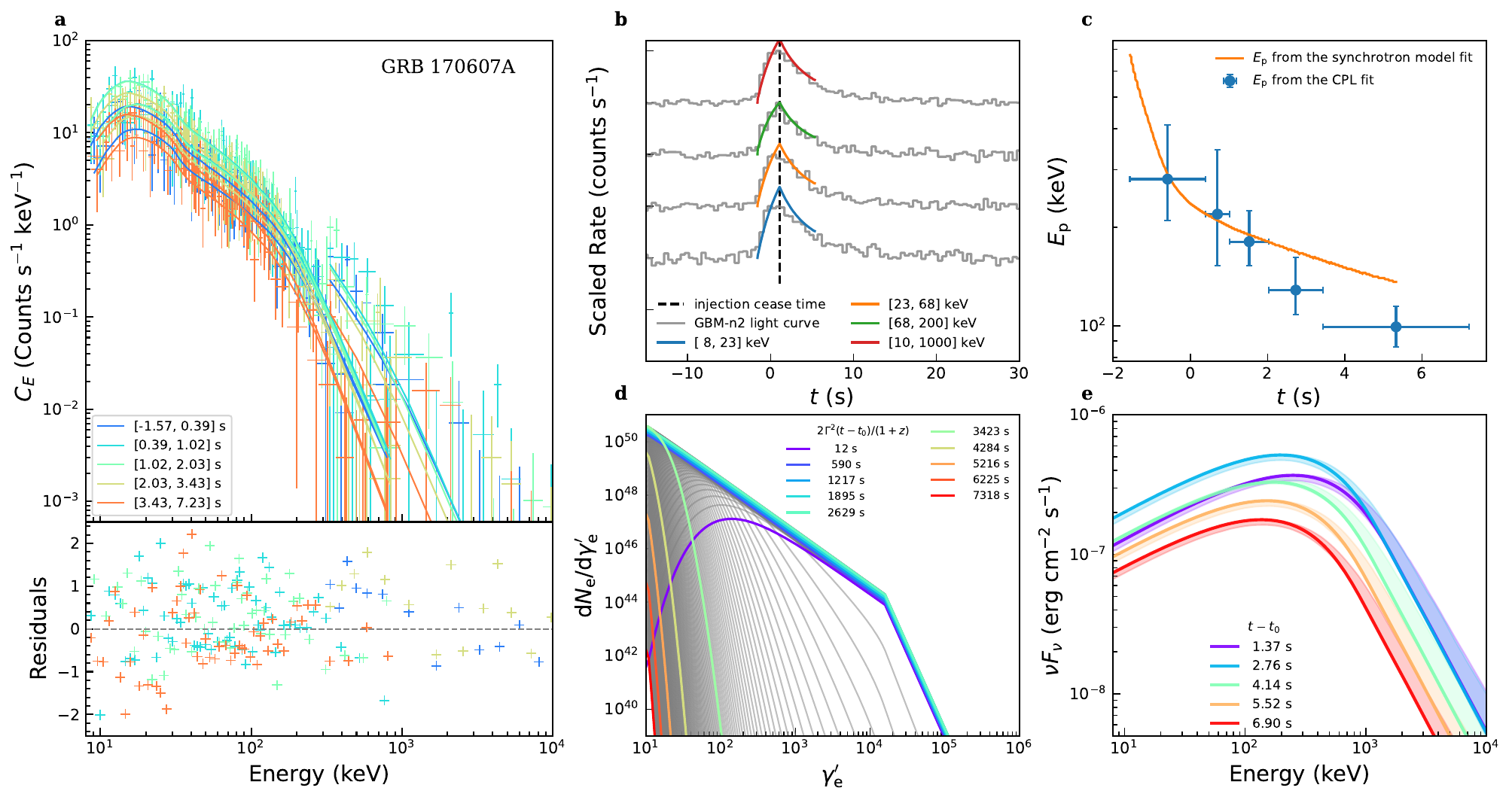}
\caption{Same as Figure \ref{fig:120326A}, but for GRB 170607A.
\label{fig:170607A}}
\end{figure}

\begin{figure}[h!]
\plotone{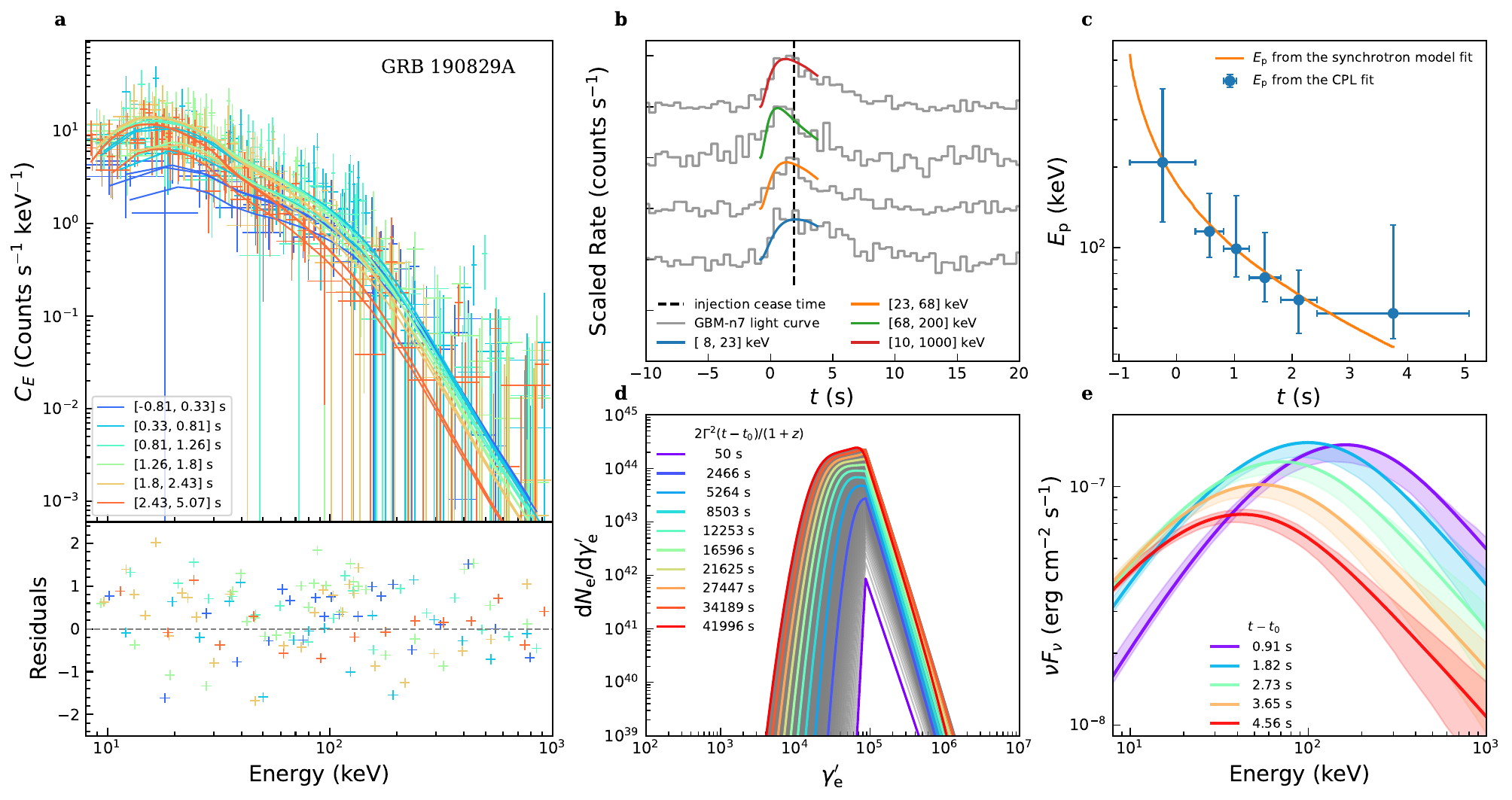}
\caption{Same as Figure \ref{fig:120326A}, but for GRB 190829A.
\label{fig:190829A}}
\end{figure}

\begin{figure}[h!]
\plotone{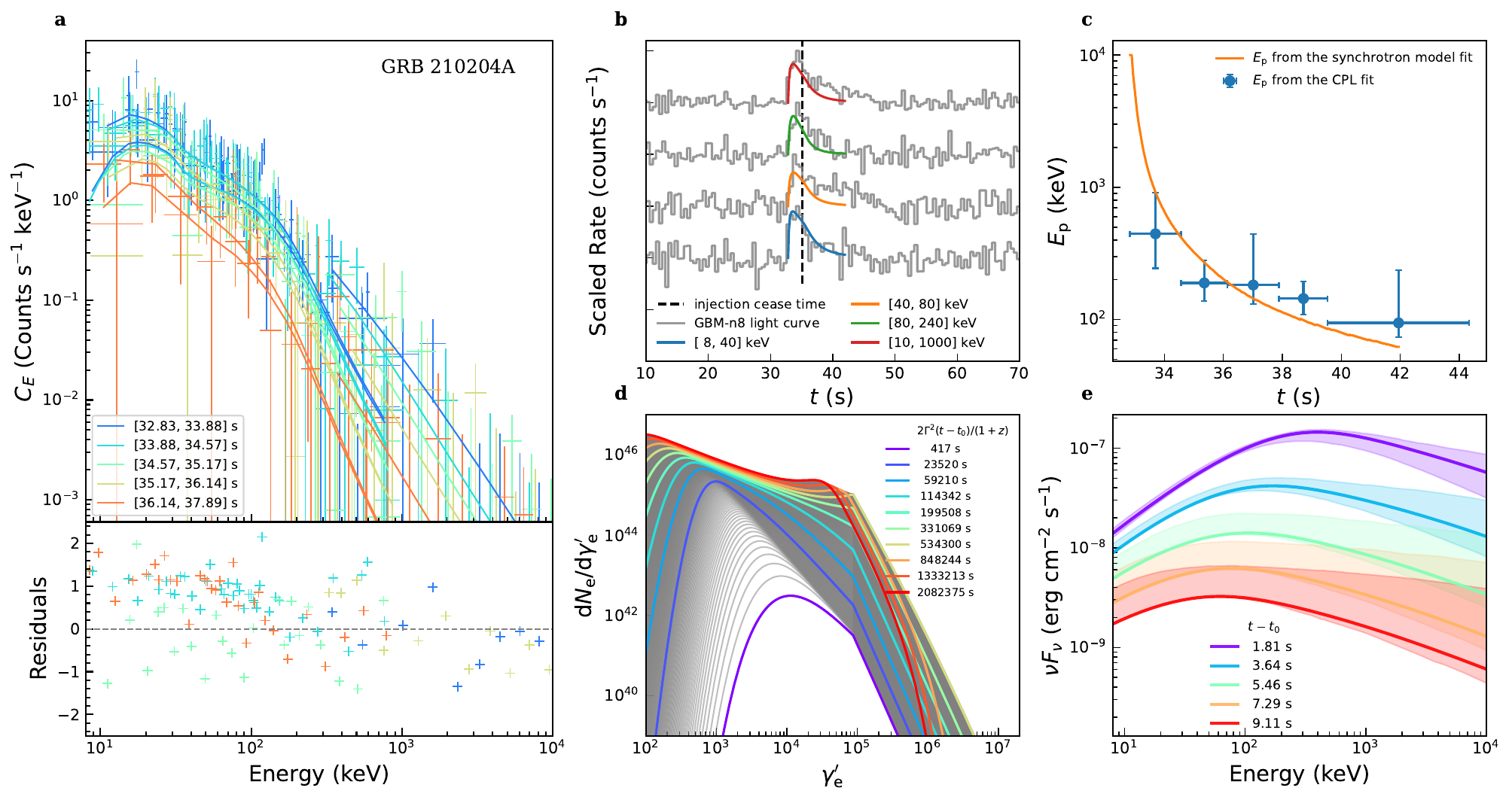}
\caption{Same as Figure \ref{fig:120326A}, but for GRB 210204A.
\label{fig:210204A}}
\end{figure}

The fitting results of the synchrotron model for our sample except GRB 120326A are presented in Figure \ref{fig:131011A}--\ref{fig:210204A}.

\begin{figure}[h!]
\plotone{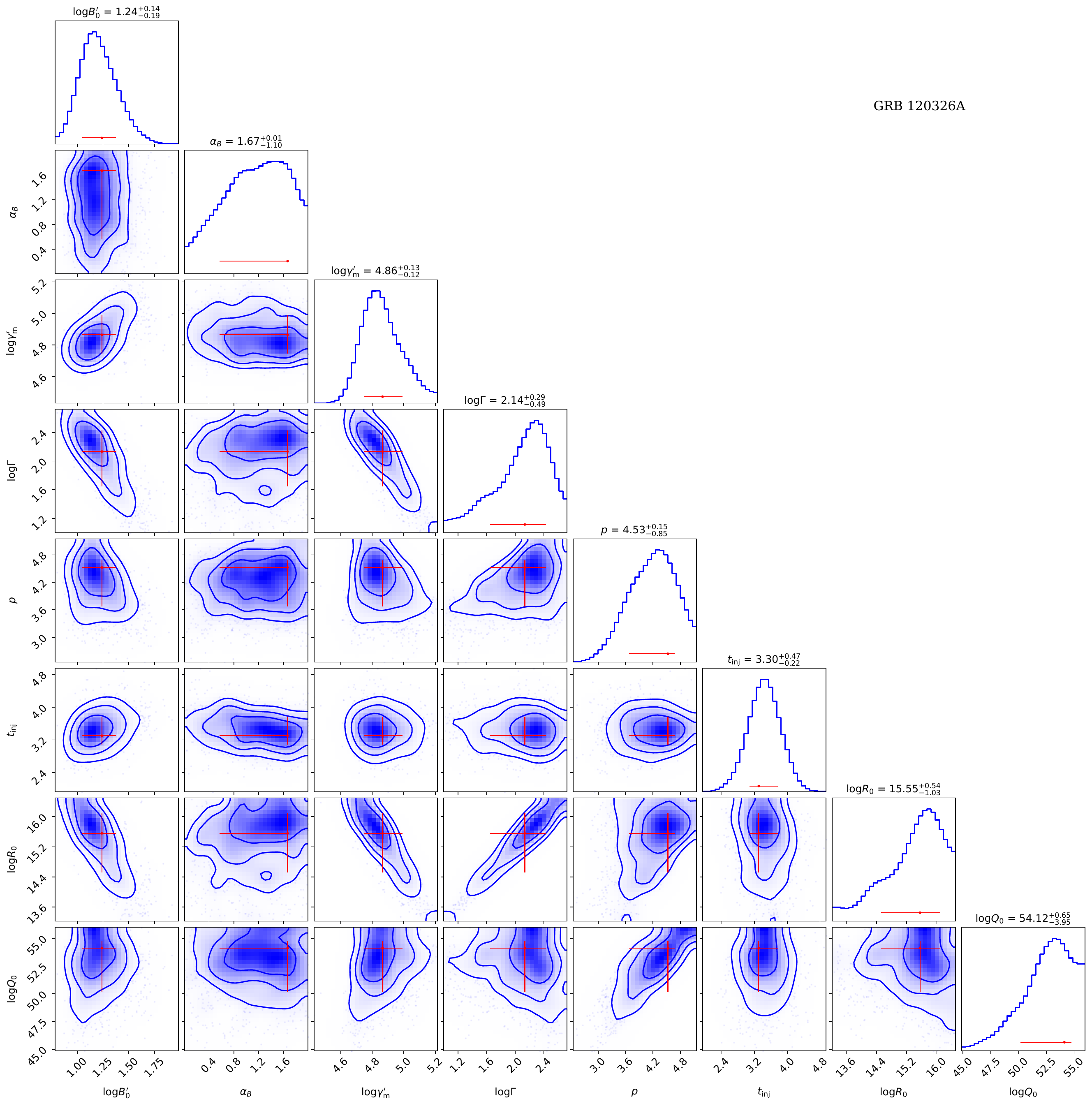}
\caption{Corner plot of the parameters' posterior probability distributions for GRB 120326A. The error bars represent $1\sigma$ uncertainties.
\label{fig:fit120326A}}
\end{figure}

\begin{figure}[h!]
\plotone{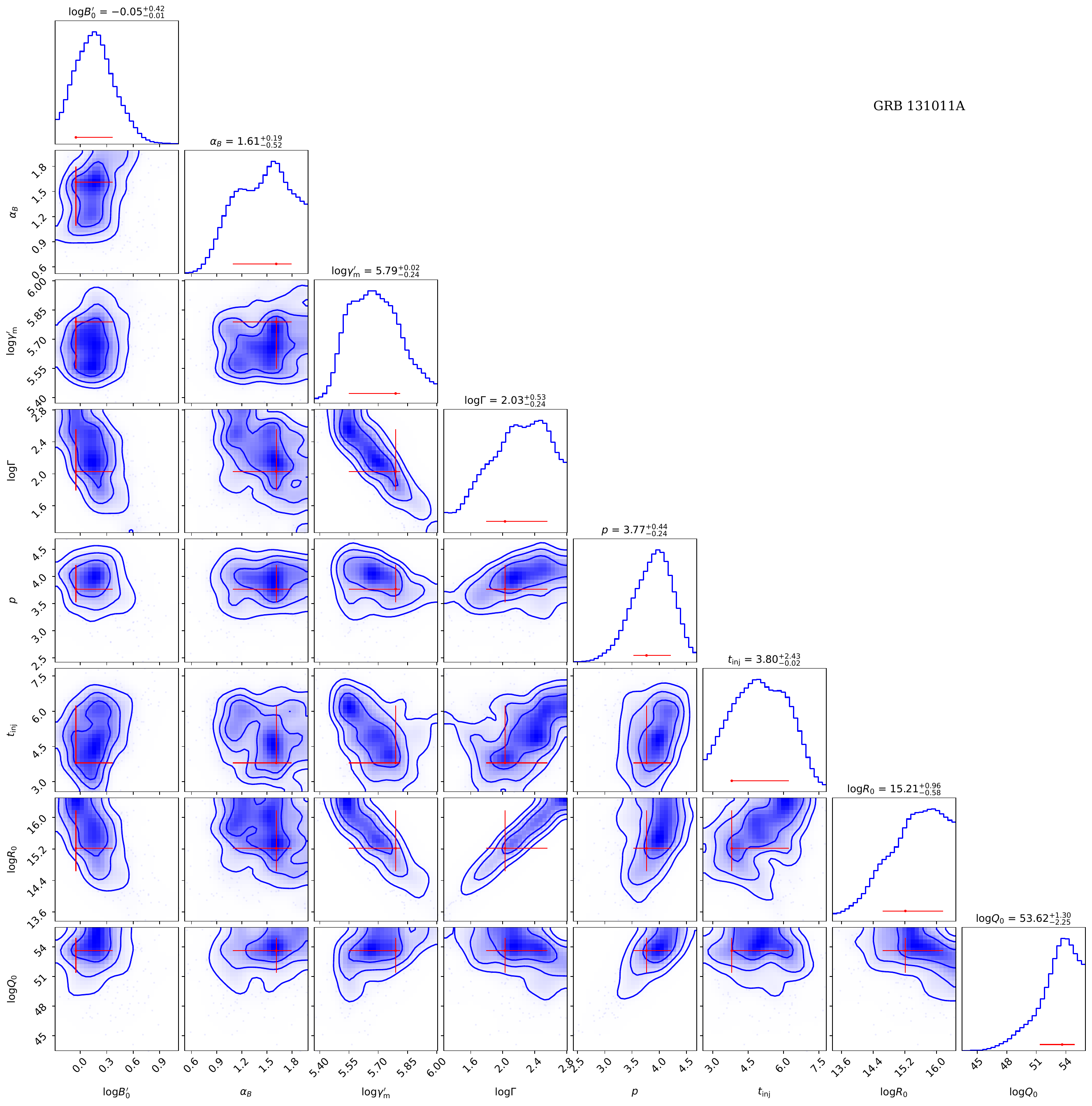}
\caption{Same as Figure \ref{fig:fit120326A}, but for GRB 131011A.
\label{fig:fit131011A}}
\end{figure}

\begin{figure}[h!]
\plotone{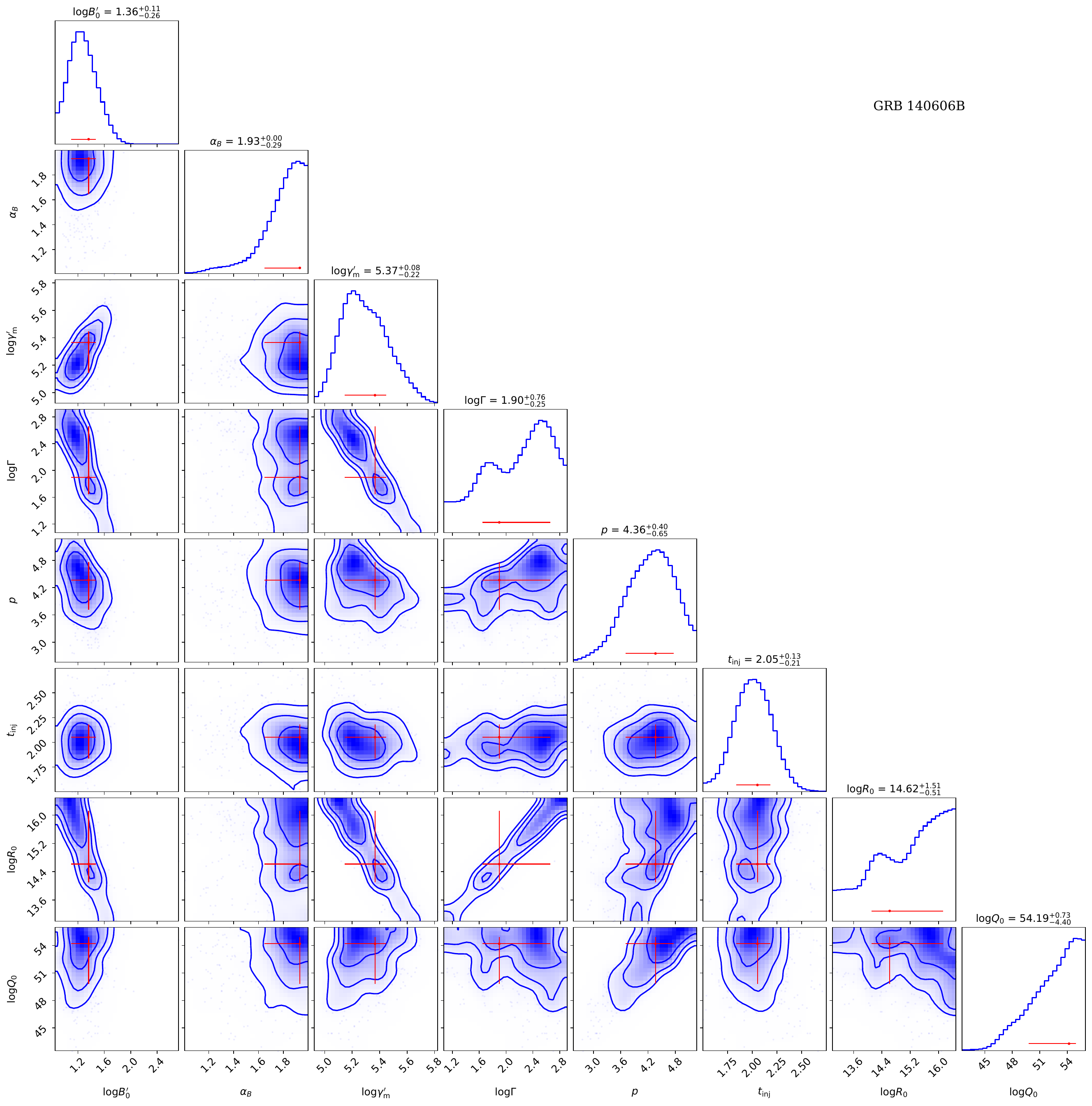}
\caption{Same as Figure \ref{fig:fit120326A}, but for GRB 140606B.
\label{fig:fit140606B}}
\end{figure}

\begin{figure}[h!]
\plotone{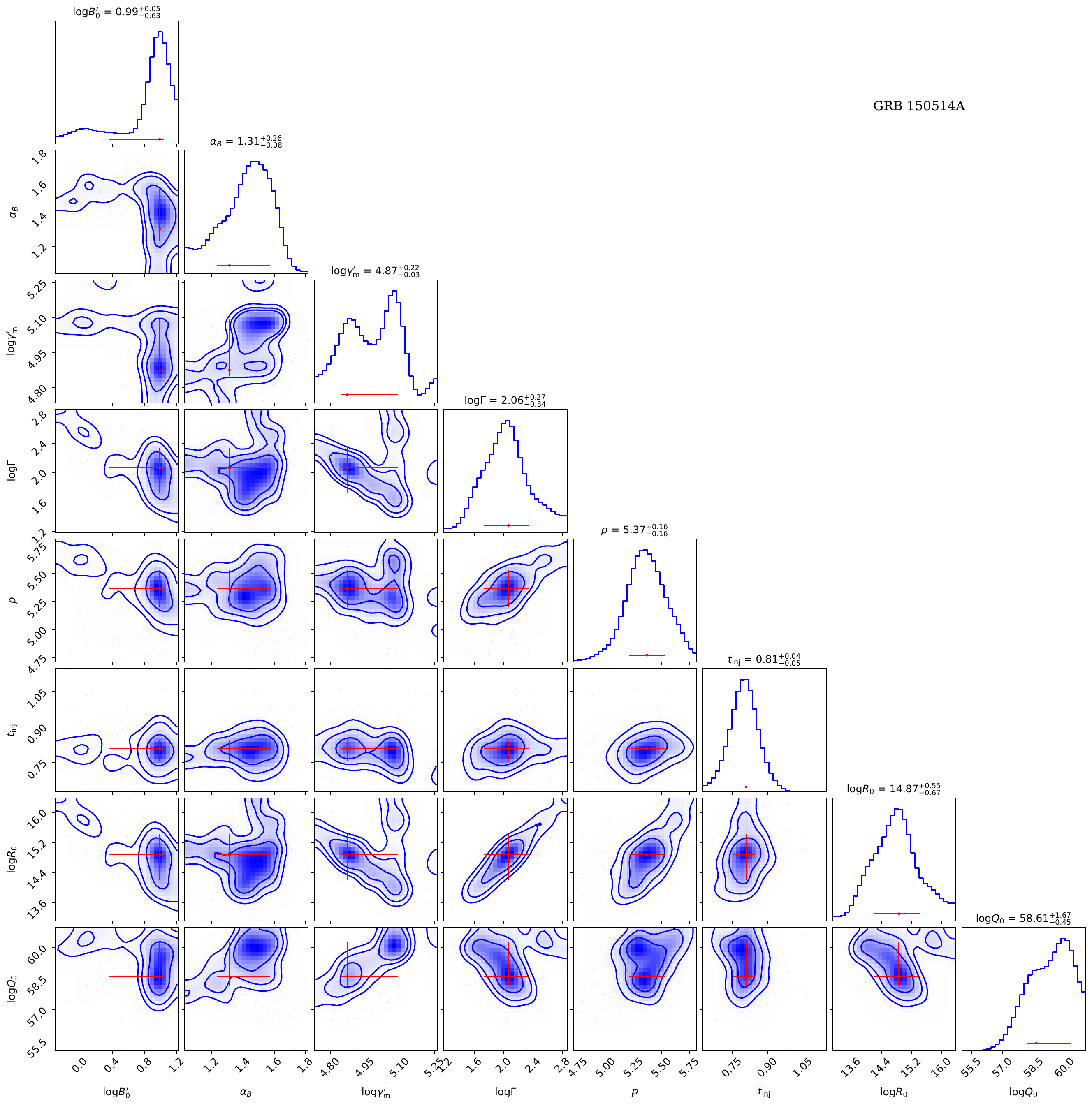}
\caption{Same as Figure \ref{fig:fit120326A}, but for GRB 150514A.
\label{fig:fit150514A}}
\end{figure}

\begin{figure}[h!]
\plotone{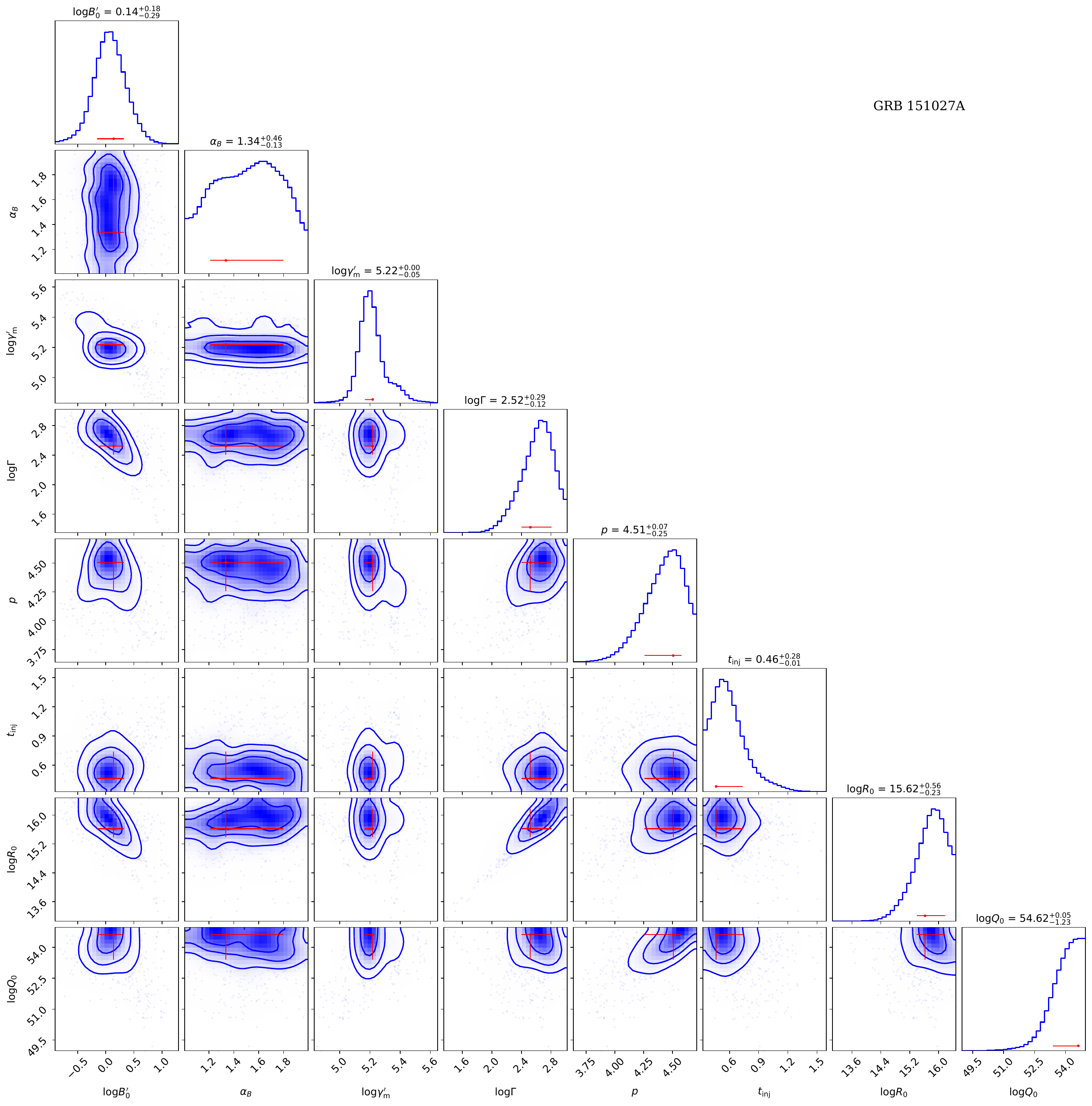}
\caption{Same as Figure \ref{fig:fit120326A}, but for GRB 151027A.
\label{fig:fit151027A}}
\end{figure}

\begin{figure}[h!]
\plotone{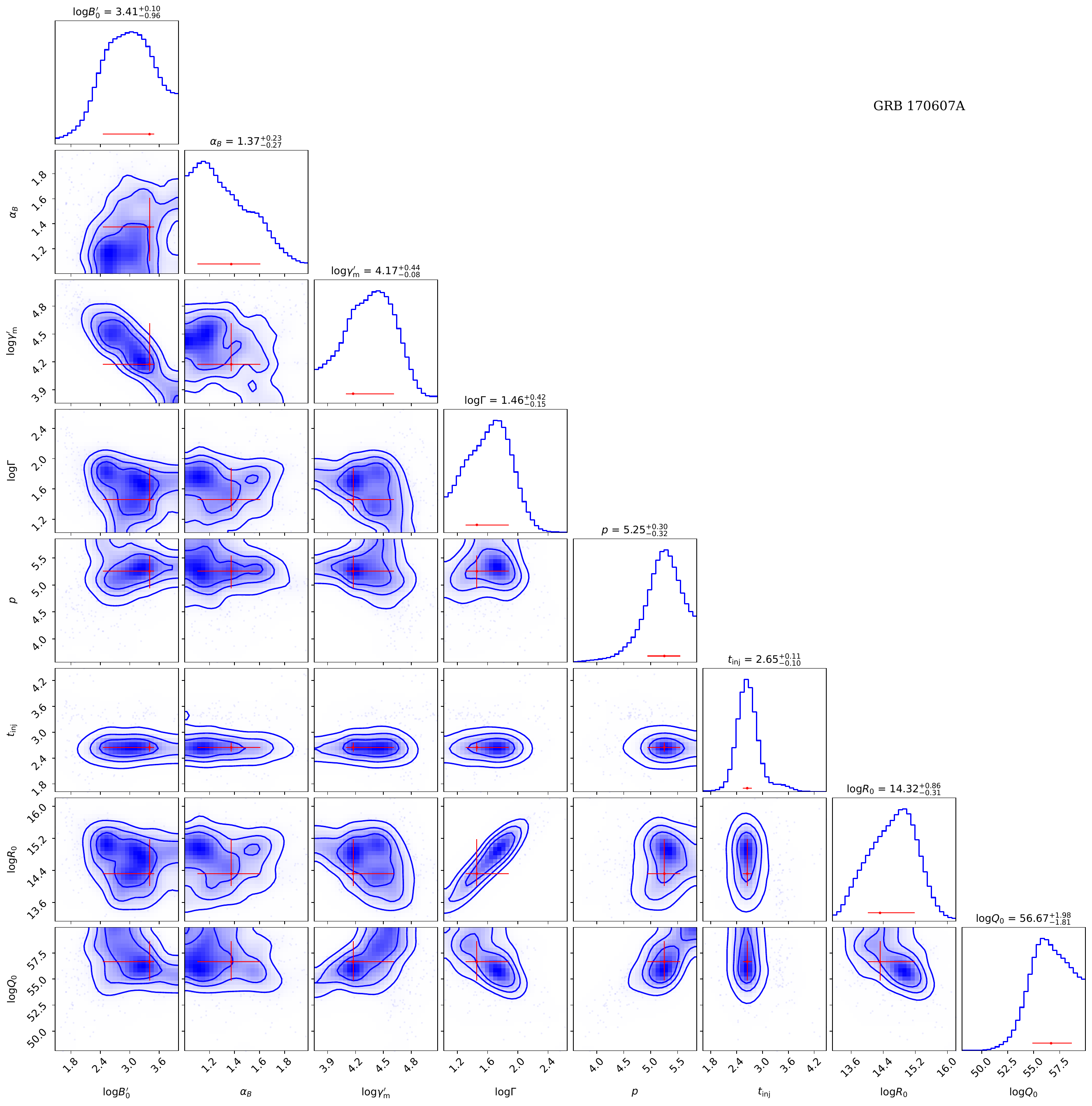}
\caption{Same as Figure \ref{fig:fit120326A}, but for GRB 170607A.
\label{fig:fit170607A}}
\end{figure}

\begin{figure}[h!]
\plotone{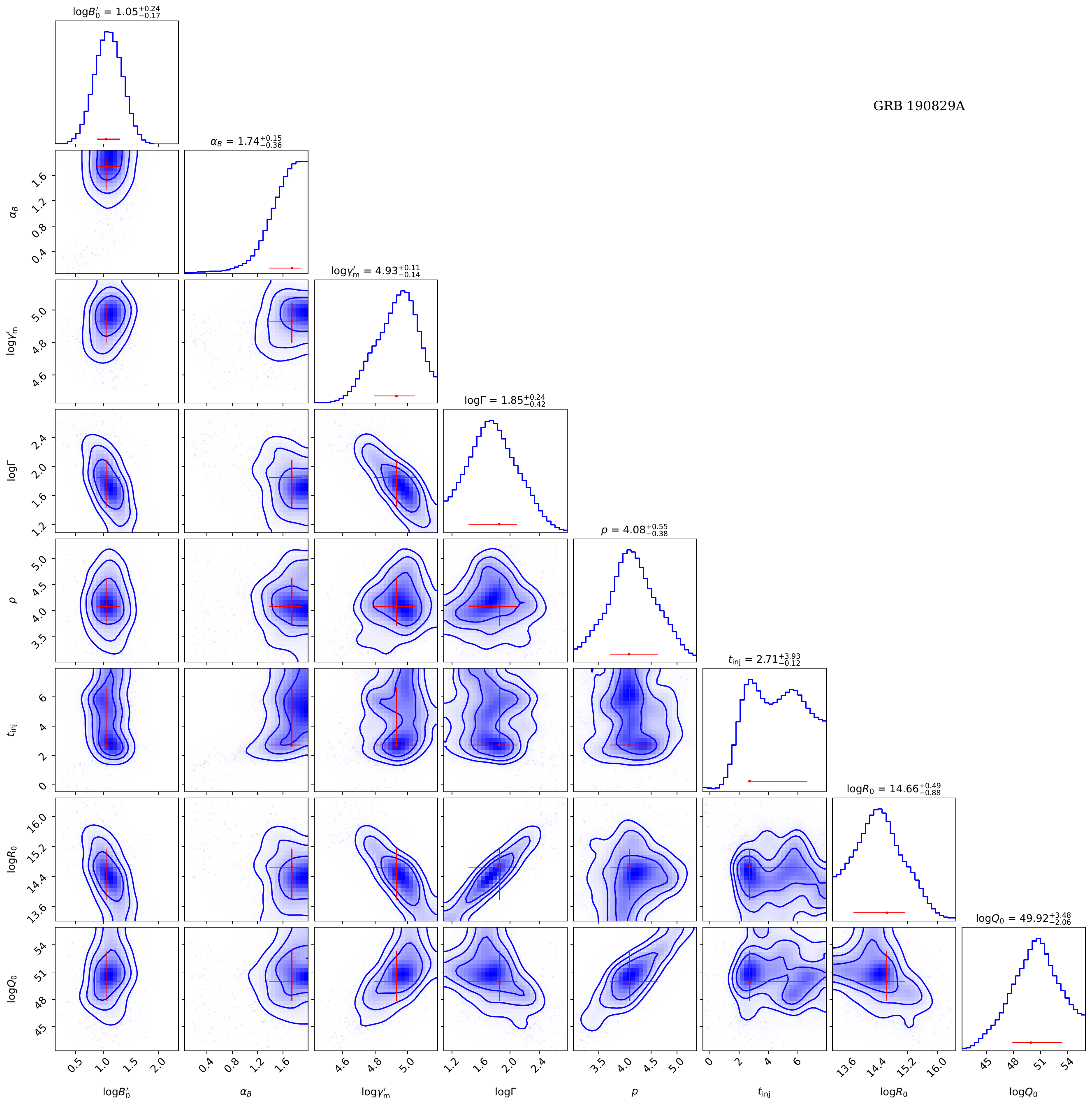}
\caption{Same as Figure \ref{fig:fit120326A}, but for GRB 190829A.
\label{fig:fit190829A}}
\end{figure}

\begin{figure}[h!]
\plotone{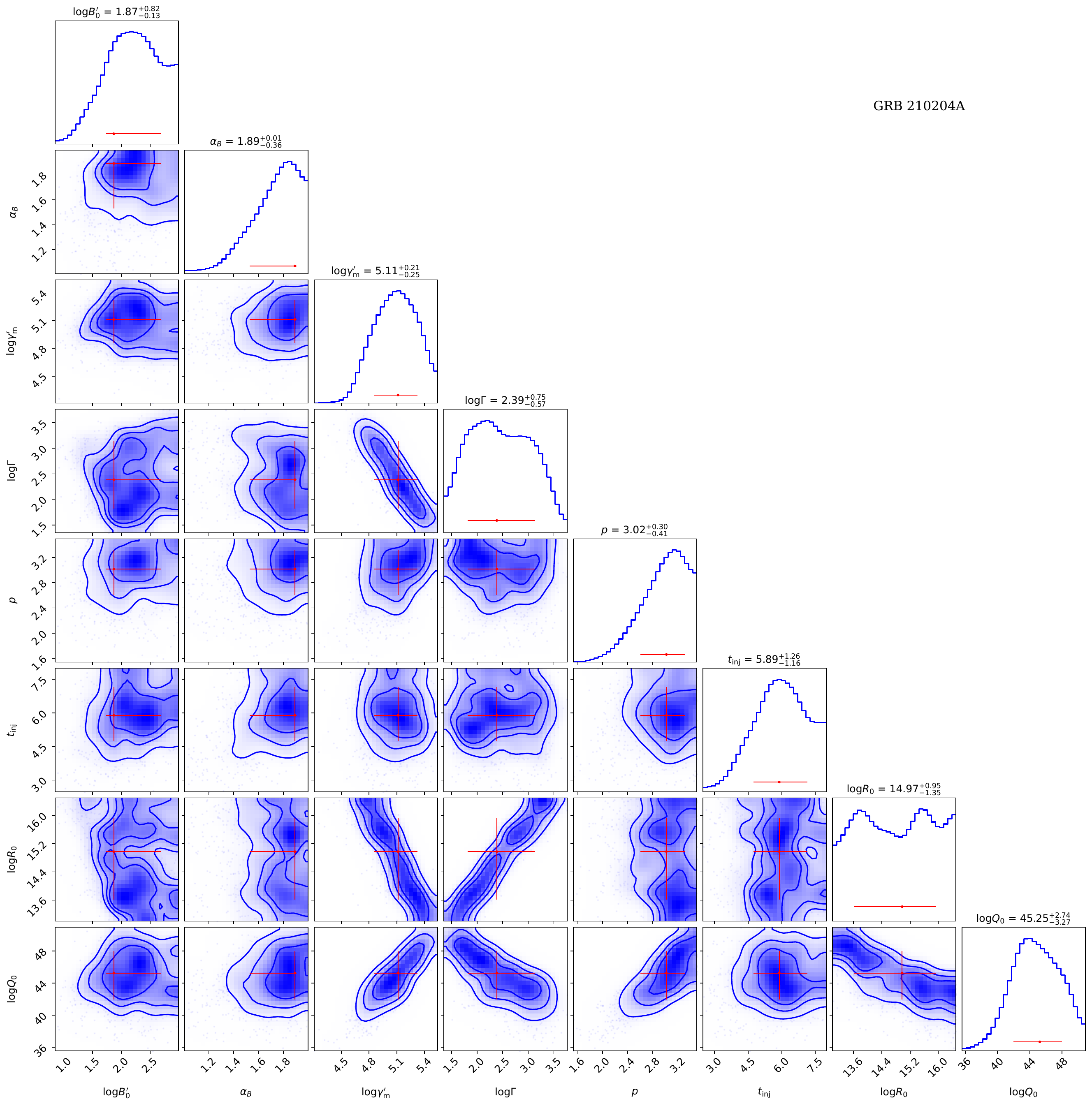}
\caption{Same as Figure \ref{fig:fit120326A}, but for GRB 210204A.
\label{fig:fit210204A}}
\end{figure}

The posterior corner plots of the fitting parameters are displayed in Figure \ref{fig:fit120326A}--\ref{fig:fit210204A}.

\begin{table*}
\centering
\small
\addtolength{\leftskip} {-3.4cm}
\setlength\tabcolsep{3pt}
\caption{Selected {\it Fermi}/GBM Detectors and Boundaries of Time Slices for Each GRB}
\label{table:Detectors}
\begin{threeparttable}
\begin{tabular}{clcllllllllll}
\toprule
GRB & detectors & $t_{\rm 0}$ & $t_{\rm 1}$ & $t_{\rm 2}$ & $t_{\rm 3}$ & $t_{\rm 4}$ & $t_{\rm 5}$ & $t_{\rm 6}$ & $t_{\rm 7}$ & $t_{\rm 8}$ & $t_{\rm 9}$ & $t_{\rm 10}$ \\ \hline
 120326A & n0,n1 & -2.51 & 0.49 & 1.49 & 2.19 & 2.89 & 7.49 & - & - & - & - & - \\
 131011A & b1,n9,na,nb & -2.99 & 1.74 & 3.56 & 5.066 & 8.61 & 12.746 & 25.01 & - & - & - & - \\
 140606B & b0,n3,n4,n8 & -0.83 & 1.345 & 1.795 & 2.34 & 3.34 & 4.43 & 5.74 & - & - & - & - \\
 150514A & b0,n3,n6,n7 & -0.46 & 0.34 & 0.60 & 0.83 & 1.05 & 1.30 & 1.55 & 1.85 & 2.25 & 3.05 & 6.05 \\
 151027A & b0,n0,n1,n3 & -0.5 & 0.58 & 1.10 & 1.60 & 2.30 & 3.50 & - & - & - & - & - \\
 170607A & b0,n2,n5 & -1.57 & 0.39 & 1.02 & 2.03 & 3.43 & 7.23 & - & - & - & - & - \\
 190829A & n6,n7,n9 & -0.81 & 0.33 & 0.81 & 1.26 & 1.80 & 2.43 & 5.07 & - & - & - & - \\
 210204A & b1,n7,n8 & 32.83 & 34.565 & 36.14 & 37.885 & 39.54 & 44.34 & - & - & - & - & - \\
\hline
\end{tabular}
\end{threeparttable}
\end{table*}

\begin{table}[h!]
\centering
\small
\addtolength{\leftskip} {-2.3cm}
\caption{Parameter Prior Bounds for the Fitting}
\label{table:fit_bounds}
\begin{threeparttable}
\begin{tabular}{ccccccccc}
\toprule
GRB & $\mathrm{log} [B^{\prime}_{\rm 0}(\mathrm{G})]$ & $\alpha_B$ & $\mathrm{log} \gamma^{\prime}_{\rm m}$ & $\mathrm{log} \Gamma$ & $p$ & $t_\mathrm{inj}(\mathrm{s})$ & $\mathrm{log} [R_\mathrm{0}(\mathrm{cm})]$ & $\mathrm{log} [Q_0({\rm cm}^{-1})] $ \\ \hline
120326A & $[-1,3]$ & $[0.0,2.0]$ & $[3,7]$ & $[1,4]$ & $[2,6]$ & $[0,8]$ & $[13,16.5]$ & $[35,56]$ \\
131011A & $[-1,3]$ & $[0.0,2.0]$ & $[3,7]$ & $[1,4]$ & $[2,6]$ & $[0,8]$ & $[13,16.5]$ & $[35,56]$ \\
140606B & $[-1,3]$ & $[1.0,2.0]$ & $[3,7]$ & $[1,4]$ & $[2,6]$ & $[0,8]$ & $[13,16.5]$ & $[35,56]$ \\
150514A & $[-1,3]$ & $[1.0,2.0]$ & $[3,7]$ & $[1,4]$ & $[2,6]$ & $[0,8]$ & $[13,16.5]$ & $[46,61]$ \\
151027A & $[-1,3]$ & $[1.0,2.0]$ & $[3,7]$ & $[1,4]$ & $[2,6]$ & $[0,8]$ & $[13,16.5]$ & $[35,55]$ \\
170607A & $[-2,4]$ & $[1.0,2.0]$ & $[3,7]$ & $[1,4]$ & $[2,6]$ & $[0,8]$ & $[13,16.5]$ & $[35,60]$ \\
190829A & $[0,3]$ & $[0.0,2.0]$ & $[3,7]$ & $[1,4]$ & $[3,7]$ & $[0,8]$ & $[13,16.5]$ & $[41,56]$ \\
210204A & $[0,3]$ & $[1.0,2.0]$ & $[3,7]$ & $[1,4]$ & $[1,3.5]$ & $[0,8]$ & $[13,16.5]$ & $[35,55]$ \\
\hline
\end{tabular}
\begin{tablenotes}
\footnotesize
\addtolength{\leftskip} {2.05cm}
\item {\it Note}. $B^{\prime}_{\rm 0}$: magnetic field strength at initial radius $R_{\rm 0}$; $\alpha_B$: magnetic field decay index; $\gamma^{\prime}_{\rm m}$: minimum injected electron energy; $\Gamma$: bulk Lorentz factor; $p$: power-law index of electron injection; $t_{\rm inj}$: cease time of electron injection ; $Q_{\rm 0}$: overall coefficient of electron injection. 
\end{tablenotes}
\end{threeparttable}
\end{table}

\end{document}